\documentclass[11pt]{article}

\usepackage{graphicx}

\usepackage[T1]{fontenc}
\usepackage[utf8]{inputenc}
\usepackage{amssymb}
\usepackage{amsfonts}
\usepackage{dsfont}
\usepackage{mathtools}
\usepackage{amsthm}
\usepackage{amsmath}
\usepackage{textcomp}
\usepackage{xcolor}
\usepackage{color,soul}
\usepackage{eurosym}
\usepackage{chngcntr}
\usepackage[numbers]{natbib}
\usepackage[english=british]{csquotes}
\usepackage{enumitem}
\usepackage{subcaption}
\usepackage{eurosym}

\counterwithout{table}{section}
\counterwithout{table}{subsection}

\usepackage{geometry}
\usepackage{blindtext}
\usepackage{hyperref}
\usepackage{cleveref}
\hypersetup{
    colorlinks=true,
    citecolor=blue,
    filecolor=blue,
    linkcolor=blue,
    urlcolor=blue
}

\def\d{\mathrm{d}}

\title{Simulating and analyzing a sparse order book: an application to intraday electricity markets}

\author{Philippe \textsc{Bergault}\footnote{Université Paris Dauphine-PSL, CEREMADE, 75775 Paris Cedex 16, bergault@ceremade.dauphine.fr.} \and Enzo \textsc{Cognéville}\footnote{EDF R\&D \& FiME, enzo.cogneville@edf.fr.} }
\date{}

\begin{document}

\maketitle

\begin{abstract}
This paper presents a novel model for simulating and analyzing sparse limit order books (LOBs), with a specific application to the European intraday electricity market. In illiquid markets, characterized by significant gaps between order levels due to sparse trading volumes, traditional LOB models often fall short. Our approach utilizes an inhomogeneous Poisson process to accurately capture the sporadic nature of order arrivals and cancellations on both the bid and ask sides of the book. By applying this model to the intraday electricity market, we gain insights into the unique microstructural behaviors and challenges of this dynamic trading environment. The results offer valuable implications for market participants, enhancing their understanding of LOB dynamics in illiquid markets. This work contributes to the broader field of market microstructure by providing a robust framework adaptable to various illiquid market settings beyond electricity trading.
\end{abstract}

\setlength\parindent{0pt}


\section{Introduction}\label{Introduction}

The study of limit order books (LOBs) is fundamental to understanding the microstructure of financial markets. A LOB records all outstanding buy and sell orders, providing a detailed snapshot of market supply and demand at various price levels. While much research has focused on liquid markets, where trading is frequent and order books are dense, illiquid markets present a different set of challenges. In illiquid markets, trading volumes are sparse, and orders are infrequent, resulting in LOBs with significant gaps between order levels. This sparsity can greatly impact market dynamics, price formation, and trading strategies.\\

Illiquid LOBs are characterized by large gaps between price levels due to the irregular and sparse nature of order placements and cancellations. The lack of continuous order flow requires models that can accurately capture the structural gaps and discontinuities within the LOB. Traditional models, which often assume a more liquid and continuously active market with a large tick-to-spread ratio, may not be suitable for these settings. Instead, there is a need for models that reflect the sparse distribution of orders and capture the distinctive characteristics of trading in illiquid markets.\\

The first empirical analysis of LOB dynamics can be dated back to the end of the 20th century, see for instance \citeauthor{biais1995empirical} \cite{biais1995empirical}, \citeauthor{luckock2001statistical} \cite{luckock2001statistical}, and \citeauthor{maslov2000simple} \cite{maslov2000simple}. With a focus on liquid stocks, \citeauthor{bouchaud2002statistical} \cite{bouchaud2002statistical}  and \citeauthor{potters2003more} \cite{potters2003more} study statistics of incoming limit order prices, the shape of the average LOB, the typical lifetime of a limit order as a function of the distance from the best price, as well as the price impact function, for which they demonstrate a logarithmic dependence on the volume traded. The price impact of specific order book events is studied in more depth in \citeauthor{eisler2012price}  \cite{eisler2012price}. A tractable LOB model based on a multiclass queueing system is proposed in \citeauthor{cont2010stochastic} \cite{cont2010stochastic}, and is then used to compute various transition probabilities of the price conditional on the state of the LOB. Several papers such as \citeauthor{cont2013price} \cite{cont2013price} or  \citeauthor{chavez2017one} \cite{chavez2017one} focus on the dynamics of the best bid and ask limits, allowing them to propose a very parsimonious model that still captures the main properties of the price process. Using an approach similar to that of \citeauthor{cont2010stochastic} \cite{cont2010stochastic}, the authors of \citeauthor{abergel2013mathematical} \cite{abergel2013mathematical} show the importance of the cancellation structure in the stability of the LOB and study the diffusive limit of the price process. In  \citeauthor{abergel2015long} \cite{abergel2015long}, they generalize their approach to incorporate Hawkes processes with exponential kernels instead of standard Poisson processes (see also the book \citeauthor{abergel2016limit} \cite{abergel2016limit}). \citeauthor{huang2015simulating} \cite{huang2015simulating} introduces the queue-reactive model, in which the intensities of the flows depend on the state of the LOB. This approach is extended in \citeauthor{huang2017ergodicity} \cite{huang2017ergodicity}, in which the authors also study the system's ergodicity and the asymptotic scaling limit of the price process. A different approach is used in \citeauthor{kelly2018markov} \cite{kelly2018markov}, where the bid and ask limit orders arrive as independent Poisson processes and the associated prices are i.i.d. random variable; in this sense, this model is the closest to the one proposed in our paper (see also \citeauthor{muni2017modelling} \cite{muni2017modelling} for a model close to the one we propose here). An agent-based model is proposed in \citeauthor{huang2019glosten} \cite{huang2019glosten}, relying on the interactions between a noise trader, an informed trader, and market makers. In the proposed framework, the whole order book shape can be derived from these simple interactions. \citeauthor{horst2018second} \cite{horst2018second} and \citeauthor{horst2017law} \cite{horst2017law} obtain fluid limits for the LOB dynamics and prove convergence of the price-volume process to coupled ODE-PDE systems. More recently, \citeauthor{cont2021stochastic} \cite{cont2021stochastic} introduced a continuous approximation of the limit order book, where the dynamics is governed by a system of stochastic partial differential equations (see \citeauthor{baldacci2024optimal} \cite{baldacci2024optimal} for an application of this framework to a problem of market making regulation).  All those papers have in a common a focus on very liquid assets, for which the bid-ask spread as well as the gaps between price levels correspond to a very small number of ticks.\\

In this paper, we propose a novel model for simulating and analyzing sparse limit order books. Our approach leverages an inhomogeneous Poisson process to model the arrival of market orders, limit orders, and cancellations on both the bid and ask sides of the LOB. This framework allows us to incorporate the sporadic nature of order flows, providing a more accurate representation of dynamics in illiquid markets where significant gaps between order levels are common.
To demonstrate the applicability and robustness of this model, we apply it to the European intraday electricity market. This market has undergone substantial transformations in recent years, driven by the increasing integration of renewable energy sources, market liberalization, and technological advancements in trading mechanisms. By modeling the LOB in this context, we aim to gain deeper insights into the unique behaviors and challenges of trading in the intraday electricity market.\\

Intraday electricity markets emerged at the beginning of the 21st century in Europe. Since 2018, the European Cross-Border Intraday Solution (XBID) connects the intraday markets of 14 countries, thus bringing more liquidity to those very illiquid markets. The behavior of those markets has been an active field of academic research since then. Using Hawkes processes, \citeauthor{favetto2019} \cite{favetto2019} and \citeauthor{graf2021} \cite{graf2021} show that the order book activity increases exponentially as time approaches maturity. The dynamics of intraday electricity prices has been studied in \citeauthor{deschatre2023common} \cite{deschatre2023common} using a common shock Poisson model that allows the authors to reproduce the correlation structure as well as the Samuelson effect, i.e. the increase of volatility as time to maturity decreases. With a different approach based on marked Hawkes processes, \citeauthor{deschatre2022electricity} \cite{deschatre2022electricity} proposes an alternative model for the price dynamics and are able to reproduce some stylized facts such as the Samuelson effect.\\

Our study contributes to the existing literature in several ways. First, it provides a comprehensive examination of LOB dynamics in illiquid markets, addressing a significant gap in market microstructure research. Second, our model offers a robust framework for simulating order flows and cancellations, adaptable to various illiquid market settings beyond electricity trading. Third, our empirical analysis and simulation results offer practical insights for market participants, including traders, regulators, and policymakers, who must navigate the complexities of intraday electricity markets.\\

The structure of this paper is as follows: Section \ref{sec:behavior} presents an empirical analysis of the behavior of LOBs on the intraday electricity market, using data provided by EPEX Spot. Section \ref{sec:math} outlines the theoretical framework of our proposed model, detailing the assumptions and mathematical foundations underpinning the inhomogeneous Poisson process. Section \ref{sec:num} discusses the simulation results, providing a comprehensive analysis of the LOB dynamics, their implications for market participants, and our suggestions for future research directions.\\

\section{Behavior of the order book}\label{sec:behavior}

This section draws heavily from the work of \citeauthor{deschatre2022electricity} \cite{deschatre2022electricity} and \citeauthor{deschatre2023common} \cite{deschatre2023common}, due to the reproducibility of their results and their straightforward approach to understanding the characteristics of the electricity market.

\subsection{Dataset}

To conduct our study, we use European electricity market data sourced from EPEX Spot, focusing specifically on the German and French markets. Each day at 3 p.m., 24 distinct markets open, each corresponding to an hour-long delivery period for the following day. These markets, termed products, end their trading periods 5 minutes before what we define as maturity. For instance, the product 18H denotes data from the delivery period starting at 6 p.m. and ending at 7 p.m. in the respective country. Although markets also exist for 30-minute and 15-minute delivery periods, we have not incorporated these into our analysis. Our study exclusively employs the 2021 dataset, encompassing daily files that comprehensively logs all market interactions across products [0H to 23H].\\

This dataset includes detailed information on market orders, limit orders, cancellations, price changes, volume changes, as well as specifics regarding the product and its corresponding delivery period. Initially, we reconstructed the supply curve for every instance of market activity by using a method close to the one described by \citeauthor{grindel2022} \cite{grindel2022}. Utilizing this reconstructed data, we can instantaneously ascertain the rank of each limit order at any given time. We consider the mid-price as the arithmetic average of the best ask price and the best bid price. Consequently, if there is no liquidity on one side (bid or ask) there is no mid-price. The tick size is 0.01\euro/MWh, and the used timestamp is accurate to one millisecond.\\

One hour before delivery, cross-border trading is not possible anymore. This change impacts liquidity and, consequently, price behavior. For this reason, we do not use data provided in this last hour.\\

\begin{figure}[h!]
\centering
\includegraphics[width=1\textwidth]{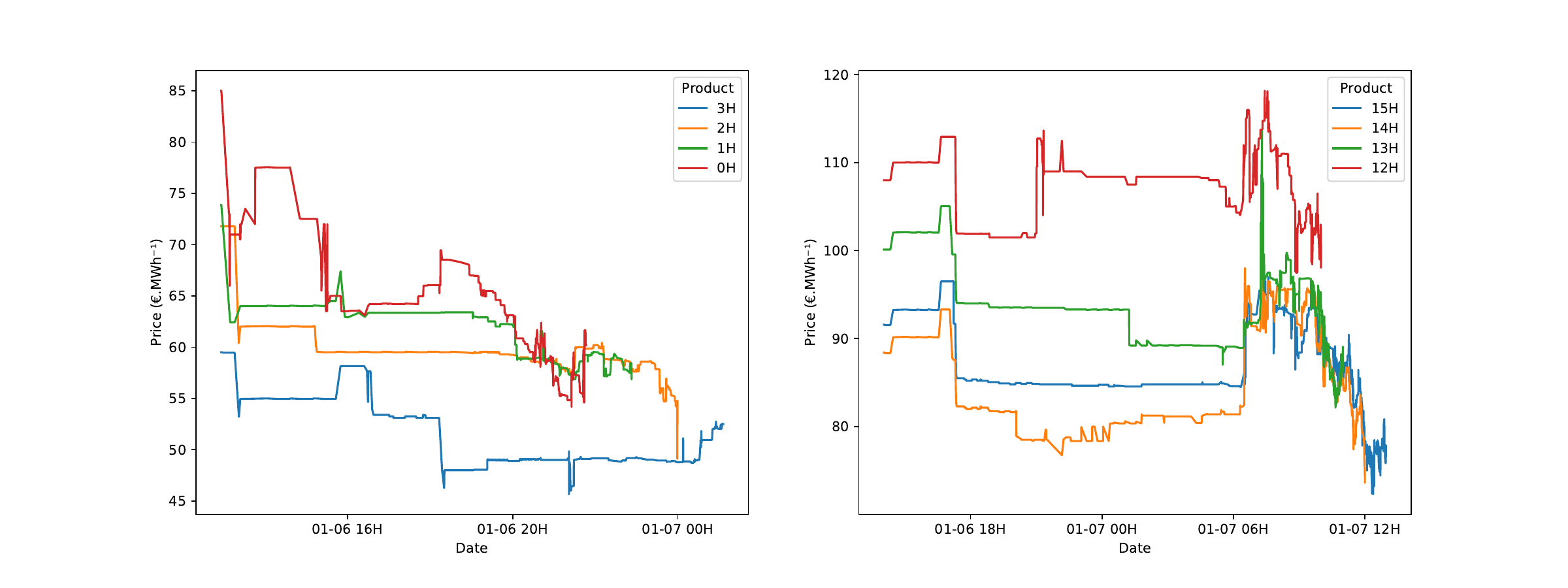}
\caption{\label{fig:midprice_evol}Intraday mid-prices evolution for the trading session on January 07, 2021}
\end{figure}

\begin{figure}[h!]
\centering
\includegraphics[width=1\textwidth]{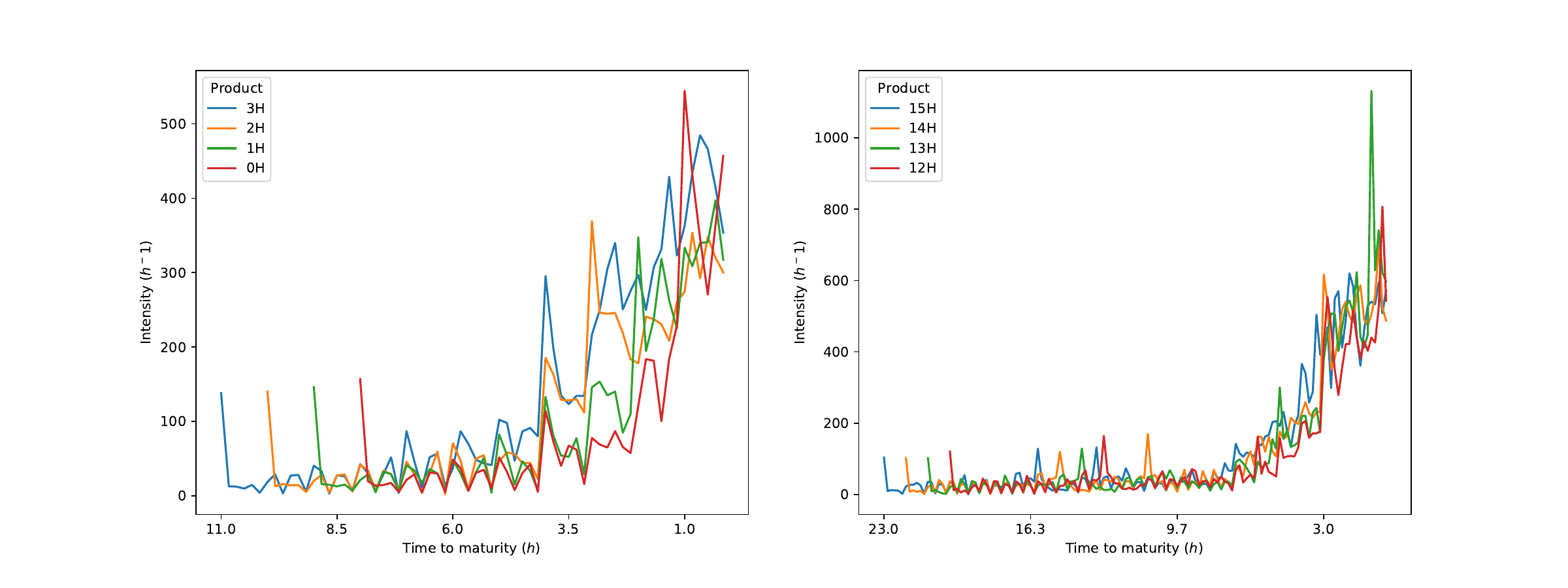}
\caption{\label{fig:cropped_intensity}Mean intensity of interactions with the market on January, 2021}
\end{figure}

To provide a comprehensive market overview, we present several statistical graphs. Figure~\ref{fig:midprice_evol} illustrates the evolution of the mid-price throughout a trading session, demonstrating a notable increase in liquidity over time. This observation is further supported by Figure~\ref{fig:cropped_intensity}, which depicts the intensity of market movements.

\subsection{Increasing intensity of actions on the market}

Numerous studies, including those by \citeauthor{favetto2019} \cite{favetto2019} and \citeauthor{graf2021} \cite{graf2021}, have documented the exponential growth of order book activity in electricity markets over trading periods. This time-dependent exponential intensity in price changes has been observed in research by \citeauthor{deschatre2022electricity} \cite{deschatre2022electricity}, as well as \citeauthor{deschatre2023common} \cite{deschatre2023common}. Our analysis of the data confirms an exponential increase in price change intensity as the time to maturity decreases. Moreover, this Samuelson effect extends beyond price changes to encompass all market interactions, as depicted in Figure~\ref{fig:cropped_intensity}. For a given
maturity, the corresponding intensity curve has been calculated from the average number of interaction changes over a trading session with a 10-minute windows. This observation was also spotted on other maturities and in the two countries considered (France and Germany).

\subsection{Sparsity and illiquidity}
\begin{figure}
  
\begin{subfigure}{.475\linewidth}
    \includegraphics[width=\linewidth]{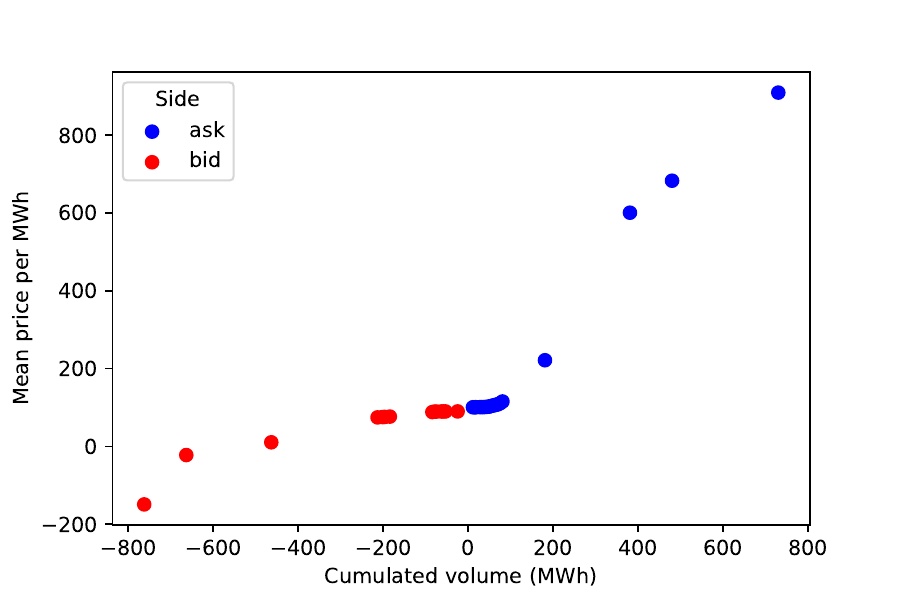}
    \caption{\label{fig:whole_liquidity_curve}Entire liquidity curve}
\end{subfigure}
\begin{subfigure}{.475\linewidth}
    \includegraphics[width=\linewidth]{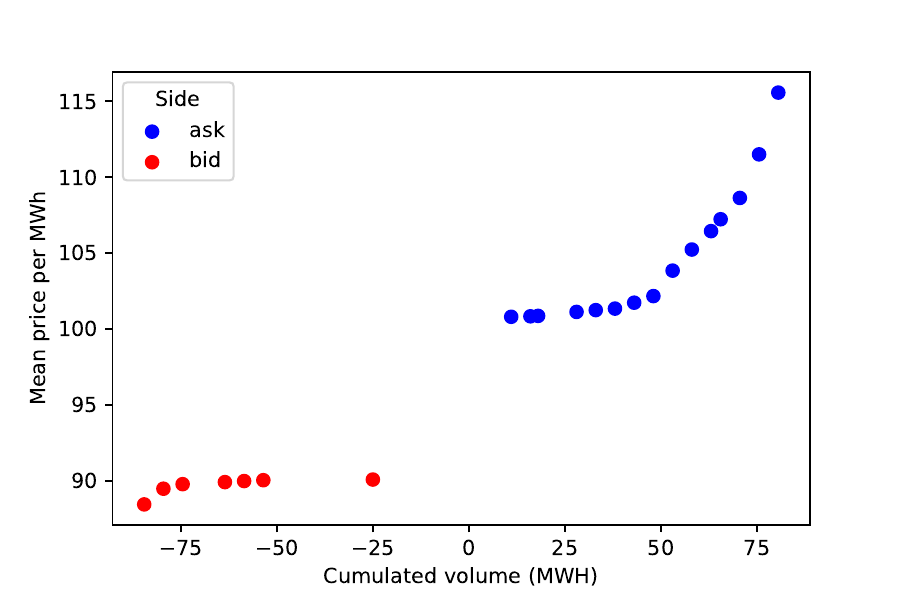}
    \caption{\label{fig:zoomed_liquidity_curve}Zoomed liquidity curve}
\end{subfigure}

\caption{\label{fig:supply_curve}Liquidity curve on January 07, 2021, for product 18H, 2 hours before maturity}
\end{figure}

\begin{figure}
\centering
\includegraphics[width=1\textwidth]{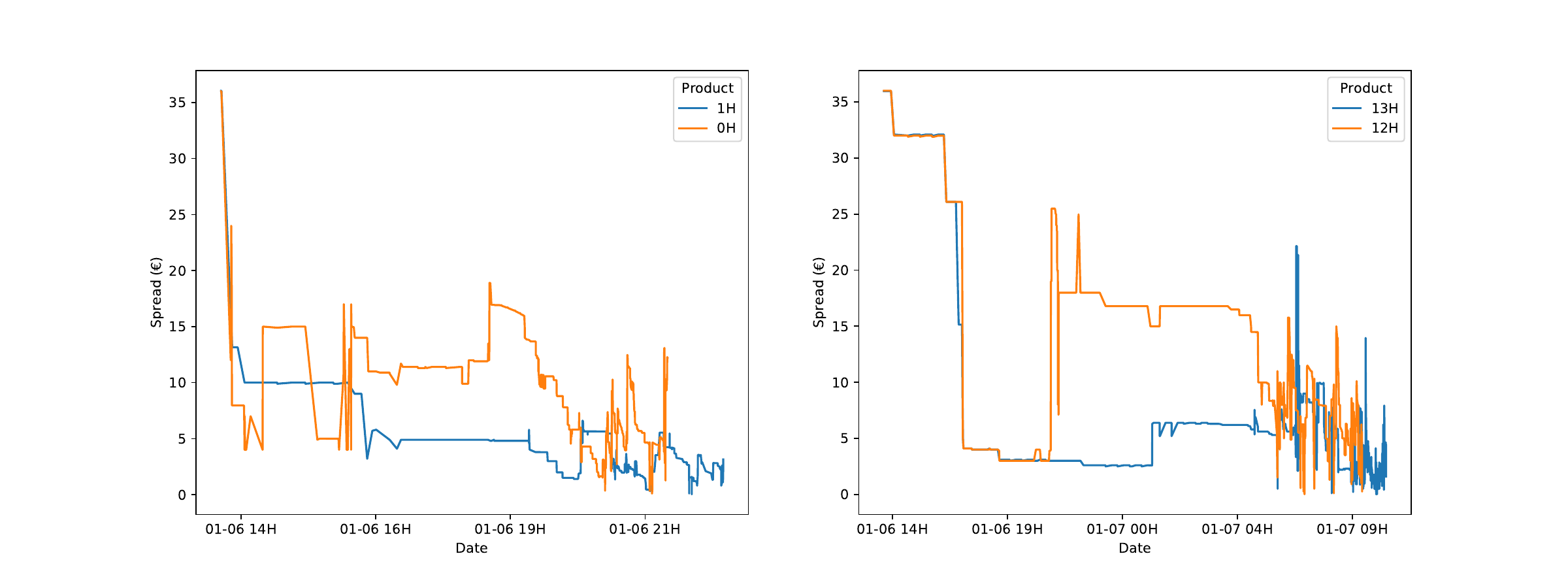}
\caption{\label{fig:spread_evol}Spread evolution on January 07, 2021}
\end{figure}

\begin{figure}[h]
\centering
\includegraphics[width=0.5\textwidth]{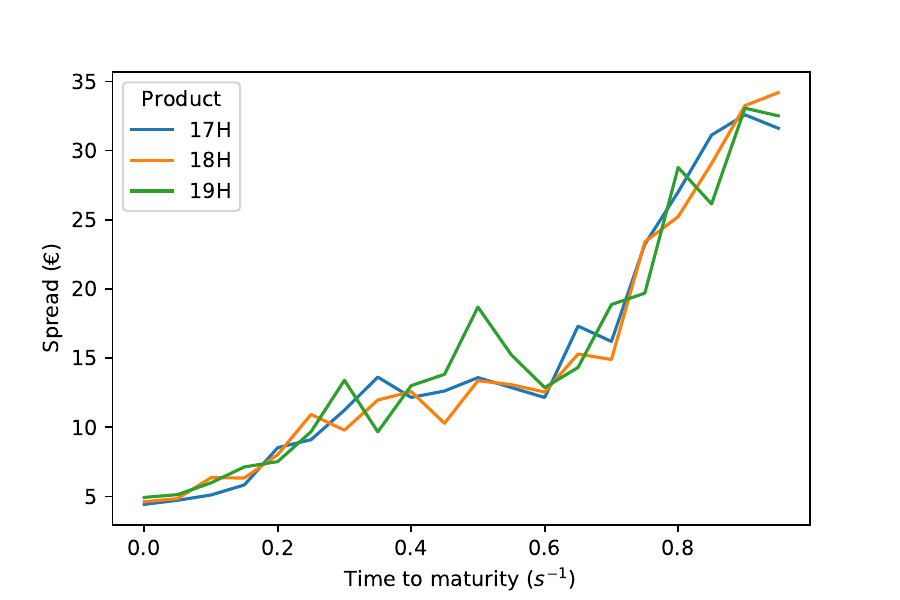}
\caption{\label{fig:mean_spread}Average spread values by normalized time to maturity on 2021}
\end{figure}

\begin{figure}[h]
\centering
\includegraphics[width=0.5\textwidth]{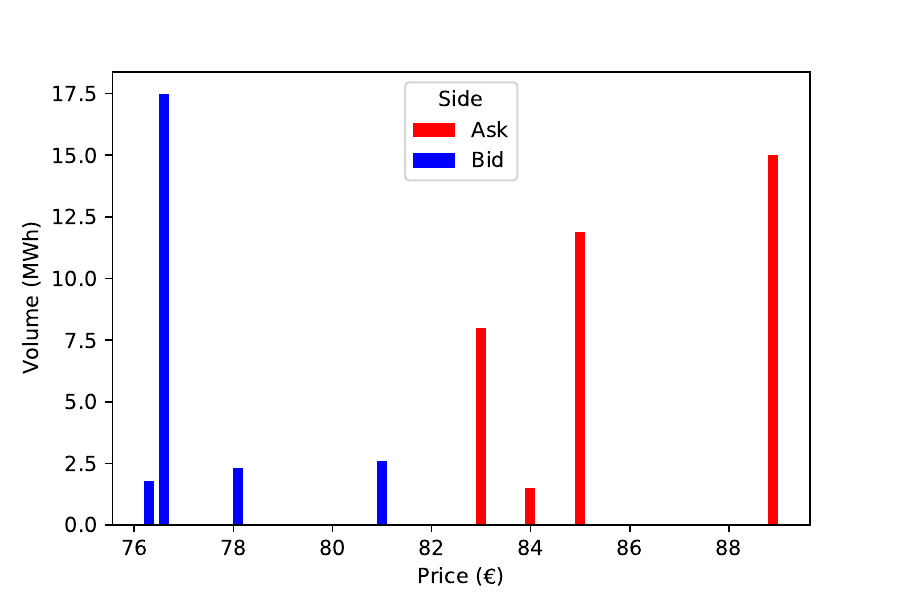}
\caption{\label{fig:sparse_order_book}Top four available limit orders on both sides of the LOB on May 5th, 2021, for product 18H, 2 hours before maturity}
\end{figure}

As observed in the last section, the liquidity of the electricity market appears to be limited. This is illustrated in Figure~\ref{fig:supply_curve}. Following the methodology of \citeauthor{blais2010} \cite{blais2010}, we made a movie of the evolution of the supply curve throughout a trading session. We saw that the supply curve become more linear at the end of session, while retaining nonlinearity for larger volumes. Concurrently, we observed a reduction in spread, as depicted in Figure~\ref{fig:spread_evol} and Figure~\ref{fig:mean_spread}. Notably, Figure~\ref{fig:zoomed_liquidity_curve} highlights a distinct nonlinearity for volumes beyond [-75; 50] MWh. Additionally, while trading volumes tend to rise towards session ends, they consistently fail to sufficiently fill most tick prices adjacent to the mid-price. This chronic liquidity deficit is further underscored by the sparse distribution of open orders at any given time $t$, as depicted in Figure~\ref{fig:sparse_order_book}. Consequently, the scarcity of available limit orders across price ticks renders the models referenced in introduction (Section \ref{Introduction}) impractical for our analysis.

\section{Mathematical model for a sparse order book}\label{sec:math}

Let $T>0$ be the time at which the market closes, and $K \in \mathbb N^\star$. We consider a $2K-$dimensional process $(\bar S_t)_t = \left(S^{-K}_t, \ldots, S^{-1}_t, S^{1}_t, \ldots, S^{K}_t \right)_t$ representing the $K$ best prices at the bid and at the ask, respectively. More precisely, for $k=1, \ldots, K$ and $t\in [0,T]$, $S^{—k}_t$ represents the $k-$th best bid price and $S^k_t$ represents the $k-$th best ask price.\\

For $t\in [0,T]$ we denote by $S^0_t$ and $\mathbf S_t$ the mid-price and the spread, respectively, at time $t$, i.e.
$$S^0_t = \frac{S^{-1}_t + S^1_t}{2}, \qquad \text{and} \qquad \mathbf S_t = S^1_t - S^{-1}_t.$$

We also introduce the 2K-dimensional process $(\bar V_t)_t = \left(V^{-K}_t, \ldots, V^{-1}_t, V^{1}_t, \ldots, V^{K}_t \right)_t$ representing the volumes associated with each of the prices above.\\

Finally, we introduce the cumulative volumes $\left( \bar{\mathcal V}_t\right)_t = \left( \mathcal V^{-K}_t, \ldots, \mathcal V^{-1}_t, \mathcal V^{1}_t, \ldots, \mathcal V^{K}_t\right)_t$ given for $k=1, \ldots, K$ by
$$\mathcal V^{-k}_t = \sum_{i=1}^k V^{-i}_t, \qquad \text{and} \qquad  \mathcal V^{k}_t = \sum_{i=1}^k V^{i}_t.$$

\subsection{Market orders}

We introduce $N^{-M}\left(\d t,\d \xi,\prod_{k=1}^{K-1}\d z_k, \prod_{k=1}^{K-1}\d \zeta_k\right)$ and $N^M\left(\d t,\d \xi,\prod_{k=1}^{K-1}\d z_k, \prod_{k=1}^{K-1}\d \zeta_k\right)$ the two marked point processes modeling the arrival of market orders at the bid and at the ask. The size of each market order is modeled by a random variable. Moreover, every time the first limit at the bid or at the ask are consumed, all the limits are shifted, and we need to introduce a new $K-$th limit. The volume at this new $K-$th limit as well as its distance to the previous $K-$th limit are modeled by a two random variables that are drawn at the time of the market order. Note that there might be several limits to introduce if several limits are consumed by the market order ; to deal with this issue, we always draw $K-1$ new limits and then only use the limits that we actually need.\\

Mathematically, we assume that $N^{—M}\left(\d t,\d \xi,\prod_{k=1}^{K-1}\d z_k, \prod_{k=1}^{K-1}\d \zeta_k\right)$ and $N^M\left(\d t,\d \xi,\prod_{k=1}^{K-1}\d z_k, \prod_{k=1}^{K-1}\d \zeta_k\right)$ have intensity kernels $\left(\nu^{-M}_t(\d \xi , \prod_{k=1}^{K-1}\d z_k , \prod_{k=1}^{K-1}\d \zeta_k)\right)_t$ and $\left(\nu^{M}_t(\d \xi , \prod_{k=1}^{K-1}\d z_k , \prod_{k=1}^{K-1}\d \zeta_k)\right)_t$ respectively given by:
$$\nu^{-M}_t\left(\d \xi , \prod_{k=1}^{K-1}\d z_k , \prod_{k=1}^{K-1}\d \zeta_k\right) = \lambda^{-M}(t, \mathbf S_{t-}) \phi^{-M}(\xi, \mathcal V^{-K}_{t-})\d \xi \prod_{k=1}^{K-1} \left(f^{-K-1}(t,z_k) \phi^{-K-1} (\zeta_k) \d z_k \d \zeta_k\right), $$
and
$$ \nu^{M}_t\left(\d \xi , \prod_{k=1}^{K-1}\d z_k , \prod_{k=1}^{K-1}\d \zeta_k\right) = \lambda^M(t, \mathbf S_{t-})\phi^{M}(\xi, \mathcal V^{K}_{t-})\d \xi \prod_{k=1}^{K-1} \left(f^{K+1}(t,z_k) \phi^{K+1} (\zeta_k) \d z_k \d \zeta_k\right),$$
where
\begin{itemize}
    \item $\lambda^{-M}$ and $\lambda^M$ are deterministic functions representing the intensity of arrival of market orders at the bid and at the ask as a function of the time and the current spread,
    \item $\phi^{-M}(\cdot, \mathcal V)$ and $\phi^{M}(\cdot, \mathcal V)$ are probability density functions with support in $(0, \mathcal V)$, representing the distribution of the volume of a market order ; more precisely, we assume that a single market order cannot consume at once an entire side of the order book, 
    \item $f^{-K-1}(t,.)$ and $f^{K+1}(t,.)$ are probability density functions representing the distribution of the distance of the new $K-$th limit to the previous $K-$th limit at the bid and at the ask, respectively,
    \item $\phi^{-K-1}$ and $\phi^{K+1}$ are probability density functions representing the distribution of the volume at the new $K-$th limit at the bid and at the ask, respectively.
\end{itemize}

\begin{figure}
    \begin{subfigure}{.5\linewidth}
        \includegraphics[width=\linewidth]{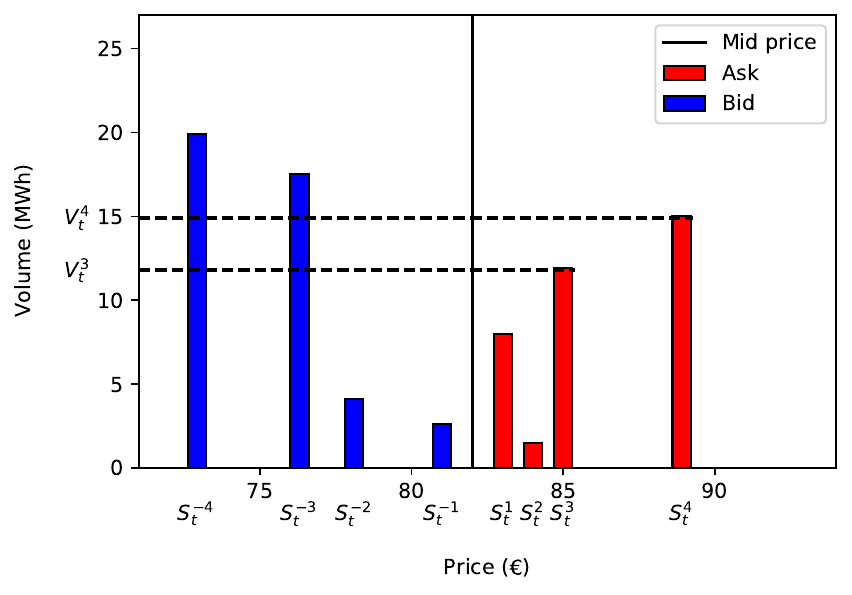}
        \caption{\label{fig:MO1}LOB at $t$}
    \end{subfigure}
    \begin{subfigure}{.5\linewidth}
        \includegraphics[width=\linewidth]{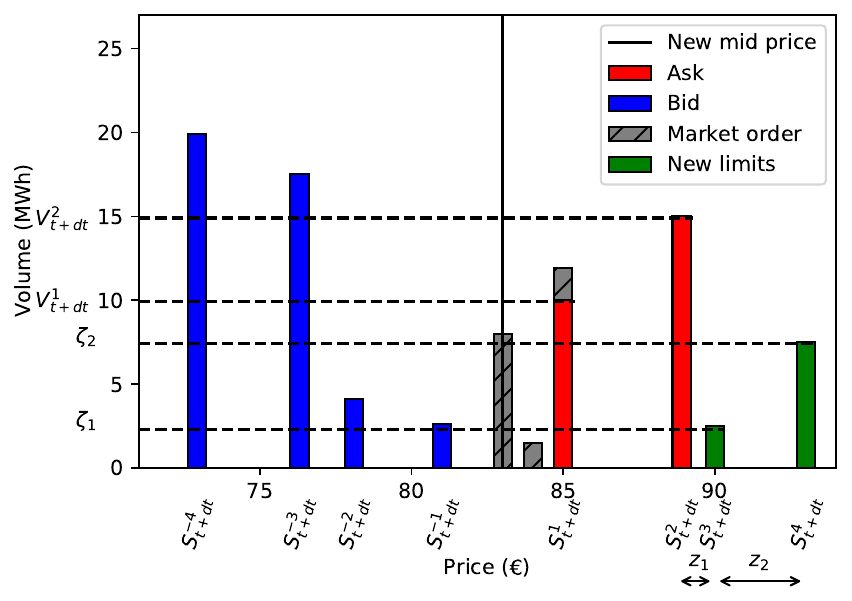}
        \caption{\label{fig:MO2}LOB at $t+dt$}
    \end{subfigure}

\caption{\label{fig:MO_LOB}Evolution of the LOB between $t$ and $t+dt$. A market order of $\xi = 12 MWh$ and generated by $\phi^M(.,\mathcal V)$ (grey area) appears and consumes more than two limits. Consequently, $S_t^3$ and $S_t^4$ are shifted to become respectively $S_{t+dt}^1$ and $S_{t+dt}^2$ at $t+dt$. Additionally, two new limits must be generated to ensure that the number of limits remains constant. The distance $z_1$ from the last existing limit $S_t^4$ is generated by $f^{K+1}(t,.)$, and its associated volume $\zeta_1$ is generated by $\phi^{K+1}$. The second limit is generated following the same process, immediately after the previous one.}
\end{figure}

\subsection{Limit orders}

There is of course a possibility that an agent inserts a new limit order between two existing limit orders, or between the first limit and the mid-price. We introduce two marked point processes $N^{-L}(\d t, \d z, \d \zeta)$ and $N^{L}(\d t, \d z, \d \zeta)$ representing the insertion of a limit order at the bid and at the ask, respectively. The intensity kernels $\left(\nu^{-L}_t(\d z , \d \zeta) \right)_t$ and $\left(\nu^{L}_t(\d z , \d \zeta) \right)_t$ of $N^{-L}(\d t, \d z, \d \zeta)$ and $N^{L}(\d t, \d z, \d \zeta)$ are respectively given by
$$\nu^{-L}_t(\d z , \d \zeta) = \lambda^{-L}(t) f^{-L}(t,z) \phi^{-L}(\zeta) \d z \d \zeta,$$
and
$$\nu^{L}_t(\d z , \d \zeta) = \lambda^{L}(t) f^{L}(t,z) \phi^{L}(\zeta) \d z \d \zeta,$$
where
\begin{itemize}
    \item $\lambda^{-L}$ and $\lambda^{L}$ are deterministic functions of time representing the intensity of insertion of a limit order at the bid and at the ask, respectively,
    \item $f^{-L}(t,.)$ is a probability density function representing the distance between the best ask $S^{1}_t$ and a newly inserted order at the bid,
    \item $f^{L}(t,.)$ is a probability density function representing the distance between the best bid $S^{-1}_t$ and a newly inserted order at the ask,
    \item $\phi^{-L}$ and $\phi^{L}$ are probability density functions representing the distribution of the volume of a new limit order at the bid and at the ask.
\end{itemize}

\begin{figure}
  
\begin{subfigure}{.475\linewidth}
    \includegraphics[width=\linewidth]{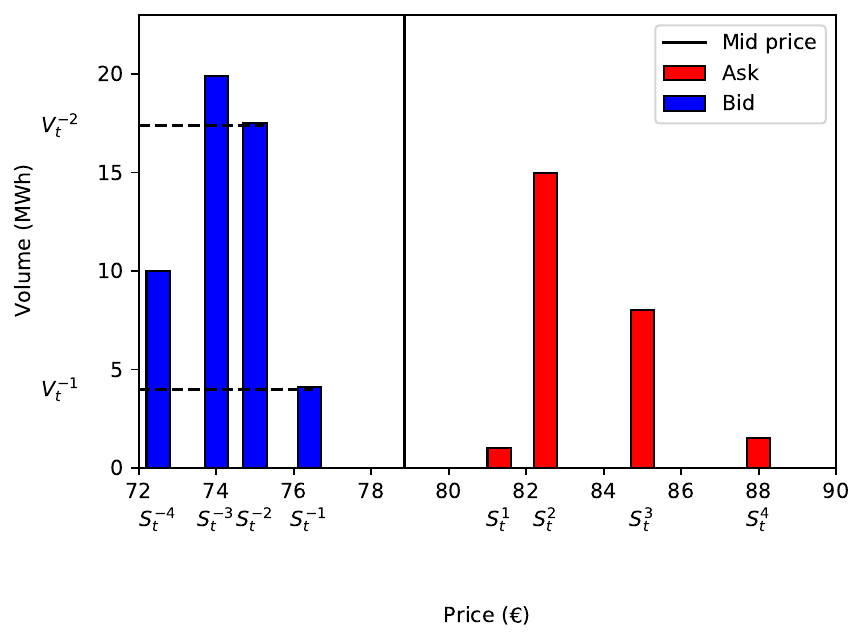}
    \caption{\label{fig:LO1}LOB at $t$}
\end{subfigure}
\begin{subfigure}{.475\linewidth}
    \includegraphics[width=\linewidth]{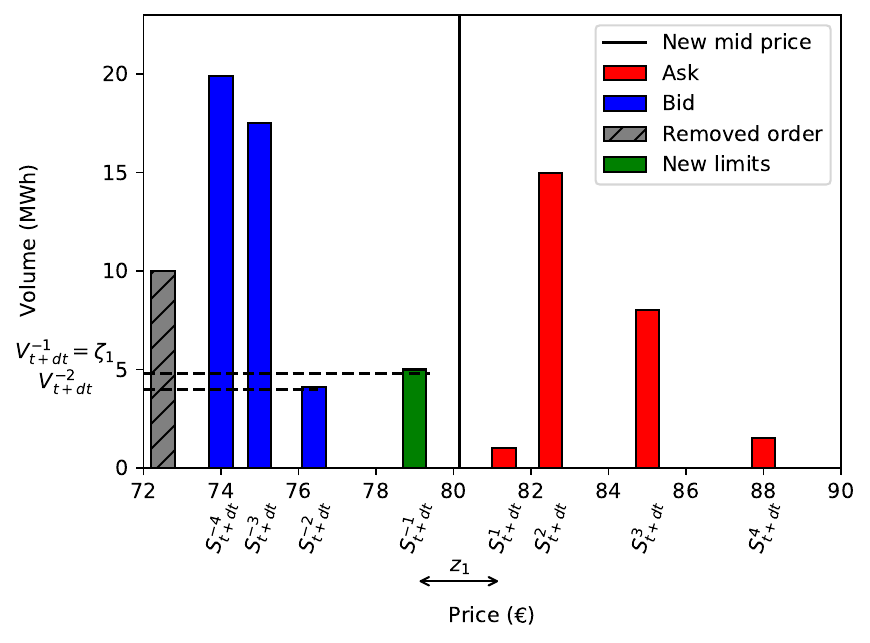}
    \caption{\label{fig:LO2}LOB at $t+dt$}
\end{subfigure}

\caption{\label{fig:LO_LOB}Evolution of the LOB between $t$ and $t+dt$. At time $t$ a new limit order arrives with a volume $\zeta$ generated by $\phi^L$ and a price that is distant from the best ask $S_t^1$ about $z_1$\euro. $z_1$ is generated by $f^{-L}(t,.)$. This new inserted limit will shift other limits.
Consequently, the last bid order must be deleted to ensure that the number of limits remains constant.}
\end{figure}

\subsection{Cancel orders}

Finally, orders can be canceled, in which case, as for market orders, all the limits are shifted, and we need to introduce a new $K-$th limit. The distance between this new $K-$th limit and the previous $K-$th limit, as well as its volume, is modeled by a random variable that is drawn at the time of the cancel. Mathematically, we introduce $K$ marked point processes at the bid $N^{-K,C}(\d t, \d z, \d \zeta), \ldots, N^{-1,C}(\d t, \d z, \d \zeta)$, and $K$ marked point processes at the ask $N^{1,C}(\d t, \d z, \d \zeta), \ldots, N^{K,C}(\d t, \d z, \d \zeta)$, where for $k=1, \ldots, K$, $N^{-k,C}(\d t, \d z, \d \zeta)$ (resp. $N^{k,C}(\d t, \d z, \d \zeta)$) represents the cancellation of the $k-$th limit order at the bid (resp. at the ask). The intensity kernel $\left(\nu^{-k, C}_t(\d z , \d \zeta) \right)_t$ of $N^{-k,C}(\d t, \d z, \d \zeta)$ is given by
$$\nu^{-k, C}_t(\d z , \d \zeta) = \lambda^{-C}\left(t\right) f^{-K-1}(t,z) \phi^{-K-1}(\zeta) \d z \d \zeta,$$
and the intensity kernel $\left(\nu^{k, C}_t(\d z , \d \zeta) \right)_t$ of $N^{k,C}(\d t, \d z, \d \zeta)$ is given by
$$\nu^{k, C}_t(\d z , \d \zeta) = \lambda^{C}\left(t\right) f^{K+1}(t,z) \phi^{K+1}(\zeta) \d z \d \zeta$$
where $\lambda^{-C}$ and $\lambda^{C}$ are deterministic functions representing the intensity of cancellation of the $k-$th limit order at the bid and at the ask, respectively. Of course, these functions could also depend on other parameters, such as the distance of order $k$ to the best order on the opposite side; however we observed no significant dependence of this type in our dataset, so we choose a simple time-dependent intensity function in order to keep the model as simple as possible.\\%We let them depend on the distance of the order to the best order on the opposite side of the book.

\begin{figure}
  
\begin{subfigure}{.452\linewidth}
    \includegraphics[width=\linewidth]{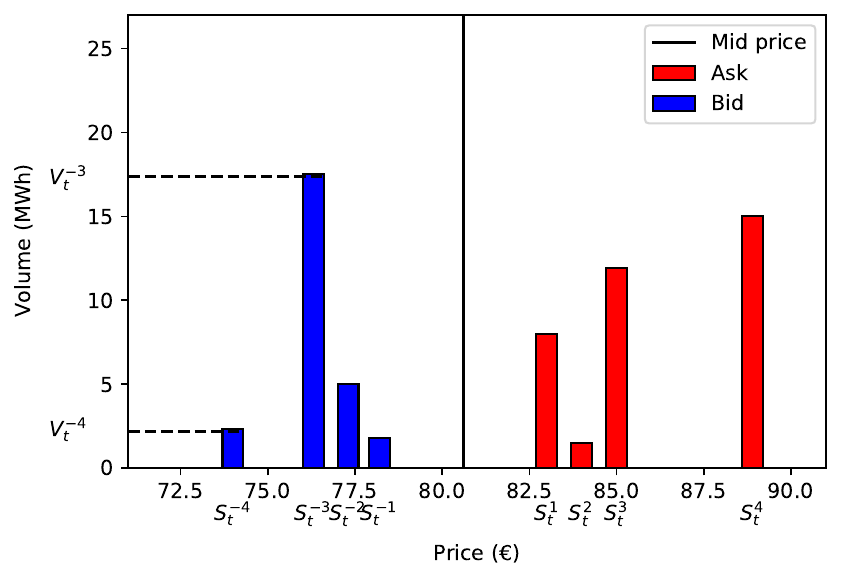}
    \caption{\label{fig:CO1}LOB at $t$}
\end{subfigure}
\begin{subfigure}{.475\linewidth}
    \includegraphics[width=\linewidth]{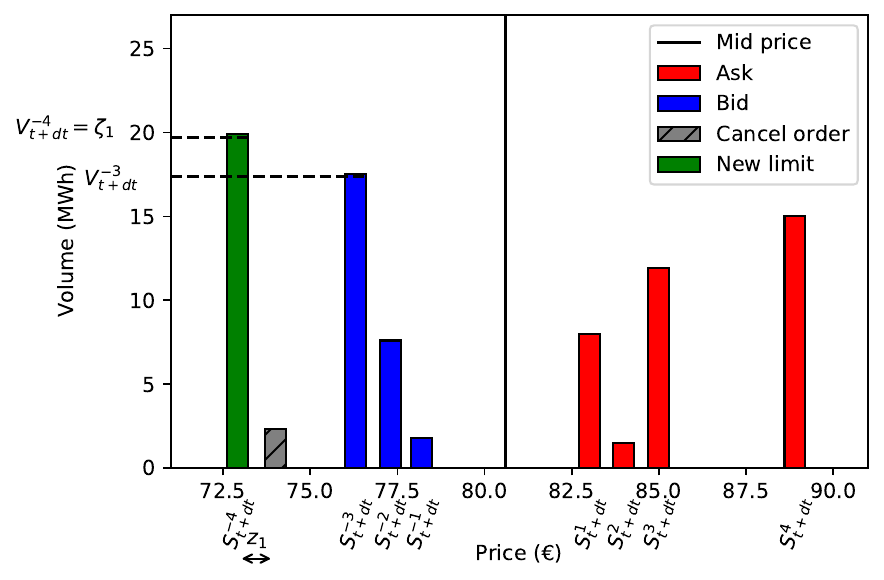}
    \caption{\label{fig:CO2}LOB at $t+dt$}
\end{subfigure}

\caption{\label{fig:CO_LOB}Evolution of the LOB between $t$ and $t+dt$. The last bid order limit is cancelled. Consequently, a new limit must be generated to ensure that the number of limits remains constant. The distance $z_1$ from the last existing limit $S_t^4$ is generated by $f^{-K-1}(t,.)$, and its associated volume $\zeta_1$ is generated by $\phi^{-K-1}$.}
\end{figure}

\subsection{Remarks}

Before moving on to the next part, a few remarks are in order. First of all, notice that we choose most of the random variables drawn at each event time to be mutually independent. This assumption is only made for the sake of simplicity, as the notations are already quite heavy, but can easily be relaxed. More generally, the model can be made more realistic by incorporating other dependence in the intensity functions, for instance on the volumes at the first limit for the intensity of market orders.\\

An important assumption in our model is that there is a single order at each limit, i.e., an agent will never insert a limit order on top of an already existing limit order. This assumption may seem very wrong in the case of highly liquid assets, but it makes sense in a sparse order book in which at each time the distance between two limit orders is very large compared to the tick size. Indeed, in this case, there is no particular reason for an agent to place a limit order at the same position as the limit order of an other agent, as she can just place it a few tick closer to the mid-price in order to be filled in priority.\\

A related point concerns the cancel orders. Indeed, contrary to the other orders, there is no volume associated with a cancel order, because we do not consider partial cancellation here, only full cancellation. Hence when an order is cancelled, the whole limit disappears. This is a small simplification made for the sake of parsimony ; a part of the effect of a partial cancellation can be taken into account in the model if we see partial cancellations in the data as a full cancellation followed by a smaller insertion at the same limit.\\

Finally, observe that, every time a limit disappear, (either because of a cancel or market order), we introduce a new limit at the end in the form of a random variable. Of course, in a real order book, this ‘‘next’’ limit generally already exists, it is just not present in our model because we only consider the $K$ first limits. This trick of inserting a new limit in order to keep the number of limits constant in our model can be seen as a simple (but reasonable nonetheless) boundary condition that yields little to no impact on the behaviour of the order book around the first limits, as long as $K$ is large enough. In this sense, if one intends to use this model to test an optimal execution algorithm, for instance, this boundary condition will have a negligible impact on the result.

\subsection{Order book dynamics}

We assume that, almost surely, there is no simultaneous jumps between all the marked point processes introduced above.\\

In order to simplify the notation (with a slight abuse), we denote respectively by $N^{-M}\left(\d t,\d \xi,\prod_{k=1}^{K-1}\d z_k\right) $ and $N^{M}\left(\d t,\d \xi,\prod_{k=1}^{K-1}\d z_k\right)$ the random measures
$$N^{-M}\left(\d t,\d \xi,\prod_{k=1}^{K-1}\d z_k\right) =  \int_{ (\zeta_1, \ldots, \zeta_{K-1}) \in \mathbb R^{K-1}_+} N^{—M}\left(\d t,\d \xi,\prod_{k=1}^{K-1}\d z_k, \prod_{k=1}^{K-1}\d \zeta_k\right)$$
and 
$$N^{M}\left(\d t,\d \xi,\prod_{k=1}^{K-1}\d z_k\right) =  \int_{ (\zeta_1, \ldots, \zeta_{K-1}) \in \mathbb R^{K-1}_+} N^{M}\left(\d t,\d \xi,\prod_{k=1}^{K-1}\d z_k, \prod_{k=1}^{K-1}\d \zeta_k\right),$$
by $N^{-M}\left(\d t,\d \xi\right) $ and $N^{M}\left(\d t,\d \xi\right)$ the random measures
$$N^{-M}\left(\d t,\d \xi\right) =  \int_{ (z_1, \ldots, z_{K-1}) \in  
\mathbb R^{K-1}_+} N^{—M}\left(\d t,\d \xi,\prod_{k=1}^{K-1}\d z_k\right)$$
and
$$N^{M}\left(\d t,\d \xi\right) =  \int_{ (z_1, \ldots, z_{K-1}) \in  
\mathbb R^{K-1}_+} N^{M}\left(\d t,\d \xi,\prod_{k=1}^{K-1}\d z_k\right),$$
by $N^{-L}(\d t, \d z) $ and $N^{L}(\d t, \d z) $ the random measures given by
$$N^{-L}(\d t, \d z) =   \int_{\zeta \in \mathbb R_+} N^{-L}(\d t, \d z, \d \zeta) \qquad \text{and} \qquad N^{L}(\d t, \d z) = \int_{\zeta \in \mathbb R_+} N^{L}(\d t, \d z, \d \zeta),$$
and by $\left(N^{-K,C}_t \right)_t, \ldots, \left(N^{-1,C}_t \right)_t$ and $\left(N^{1,C}_t \right)_t, \ldots, \left(N^{K,C}_t \right)_t$ the processes given by
$$N^{-k,C}_t = \int_0^t \int_{z \in \mathbb R_+}\int_{\zeta \in \mathbb R_+} N^{-k,C}(\d s, \d z, \d \zeta) \qquad \text{and} \qquad N^{k,C}_t = \int_0^t  \int_{z \in \mathbb R_+}\int_{\zeta \in \mathbb R_+} N^{k,C}(\d s, \d z, \d \zeta)$$
for $k=1, \ldots, K$.\\

\subsubsection{Prices dynamics}

We can now write the dynamics of the vector process $(\bar S_t)_t$. For $k=1$, we have
\begin{align*}
    \d S^{-1}_t =&\int_{\xi \in \mathbb R_+} \sum_{i=1}^{K-1} \left(S^{-i-1}_{t-} - S^{-i}_{t-} \right) \mathds{1}_{\{ \mathcal V^{-i}_{t-} \le \xi \}} N^{-M}(\d t, \d \xi) + \left(S^{-2}_{t-} - S^{-1}_{t-} \right) \d N^{-1,C}_t\\
    & + \int_{z \in \mathbb R_+} \left(S^1_{t-} - z - S^{-1}_{t-} \right) \mathds 1_{\left\{ S^1_{t-} - z \in \left[ S^{-1}_{t-},  S^{1}_{t-}\right[ \right\}} N^{-L}(\d t, \d z),
\end{align*}
and similarly
\begin{align*}
    \d S^{1}_t =&\int_{\xi \in \mathbb R_+} \sum_{i=1}^{K-1 }\left(S^{i+1}_{t-} - S^{i}_{t-} \right)\mathds{1}_{\{ \mathcal V^{i}_{t-} \le \xi \}}  N^{M}(\d t, \d \xi) + \left(S^{2}_{t-} - S^{1}_{t-} \right) \d N^{1,C}_t\\
    &+  \int_{z \in \mathbb R_+} \left(S^{-1}_{t-} + z - S^1_{t-} \right) \mathds 1_{\left\{ S^{-1}_{t-} + z \in \left] S^{-1}_{t-},  S^{1}_{t-}\right]\right\}} N^{L}(\d t, \d z).
\end{align*}
In both these equations, the first term corresponds to the arrival of a market order. If the size of the market order exceeds the cumulative volume of the first $k$ limits, then the first limit is shifted accordingly: the new first limit is the previous $(k+1)$-th limit. The second term corresponds to the cancellation of the order at the first limit: in that case, the new first limit is the previous second limit. Cancellation of orders at other limits do not impact the first limit. Finally, the last term corresponds to the arrival of a limit order within the current bid-ask spread: in that case, the new first limit is this new limit order.\\

For $1<k<K$, we have
\begin{align*}
    \d S^{-k}_t =&\int_{\xi \in \mathbb R_+} \int_{(z_1, \ldots, z_{K-1}) \in \mathbb R^{K-1}_+ } \Bigg( \mathds{1}_{\{  \xi \ge \mathcal V^{-K+k-1}_{t-} \}} \bigg(S^{-K}_{t-} - \sum_{j=1}^{K-k+1 }z_{j} -S^{-k}_{t-}\bigg)\\
    &\qquad+ \sum_{i=1}^{K-k} \left(S^{-k-i}_{t-} - S^{-k-i+1}_{t-} \right) \mathds{1}_{\{ \mathcal V^{-i}_{t-} \le \xi < \mathcal V^{-K+k-1}_{t-} \}} \Bigg) N^{-M}\left(\d t, \d \xi, \prod_{k=1}^{K-1}\d z_k\right)\\
    &+ \int_{z \in \mathbb R_+} \left(S^1_{t-} - z - S^{-k}_{t-} \right) \mathds 1_{\left\{ S^1_{t-} - z \in \left[ S^{-k}_{t-},  S^{-k+1}_{t-}\right[ \right\}} N^{-L}(\d t, \d z)\\
    &+  \int_{z \in \mathbb R_+} \left( S^{-k+1}_{t-} - S^{-k}_{t-}\right) \mathds 1_{\left\{ S^1_{t-} - z > S^{-k+1}_{t-} \right\}} N^{-L}(\d t, \d z) + \sum_{l=1}^k \left(S^{-k-1}_{t-} - S^{-k}_{t-} \right) \d N^{-l,C}_t,
\end{align*}
and similarly
\begin{align*}
    \d S^{k}_t =&\int_{\xi \in \mathbb R_+} \int_{(z_1, \ldots, z_{K-1}) \in \mathbb R^{K-1}_+ } \Bigg( \mathds{1}_{\{  \xi \ge \mathcal V^{K-k+1}_{t-} \}} \bigg(S^{K}_{t-} + \sum_{j=1}^{K-k+1 }z_{j} -S^{k}_{t-}\bigg)\\
    &\qquad+ \sum_{i=1}^{K-k} \left(S^{k+i}_{t-} - S^{k+i-1}_{t-} \right) \mathds{1}_{\{ \mathcal V^{i}_{t-} \le \xi < \mathcal V^{K-k+1}_{t-} \}} \Bigg) N^{M}\left(\d t, \d \xi, \prod_{k=1}^{K-1}\d z_k\right)\\
    & + \int_{z \in \mathbb R_+} \left(S^{-1}_{t-} + z  - S^k_t \right) \mathds 1_{\left\{ S^{-1}_{t-} + z \in \left] S^{k-1}_{t-},  S^{k}_{t-}\right]\right\}} N^{L}(\d t, \d z)\\
    &+\int_{z \in \mathbb R_+} \left( S^{k-1}_{t-} - S^{k}_{t-}\right) \mathds 1_{\left\{ S^{-1}_{t-} + z < S^{k-1}_{t-} \right\}} N^{L}(\d t, \d z) + \sum_{l=1}^k \left(S^{k+1}_{t-} - S^{k}_{t-} \right) \d N^{l,C}_t.
\end{align*}
In both these equations, the first term corresponds to the arrival of a market order. It can be decomposed into two part: the first part deals with the case where the market order consumes $K-k+1$ limits (or more), in which case new limits have to be introduce after the (previous) last limit, in the form of random variables; the second part deals with the case where the market order only consumes a small number of limits, so that the $k$-th limit can be replaced by an already existing limit and does not have to be drawn as a random variable.\\

The second term corresponds to the arrival of a limit order between the $(k-1)$-th and the $k$-th limit: in that case, the new $k$-th limit is this new limit order. The third term deals with limit orders arriving before the $(k-1)$-th limit, in which case the $k$-th limit is simply shifted.\\

The last term corresponds to the cancellation of orders before the $k$-th limit: in that case, the new $k$-th limit is the previous $(k+1)$-th limit. Cancellation of orders at limits above $k$ do not impact the $k$-th limit. \\

Finally, for $k=K$ we have
\begin{align*}
    \d S^{-K}_t =&-\int_{\xi \in \mathbb R_+} \int_{(z_1, \ldots, z_{K-1}) \in \mathbb R^{K-1}_+ } \sum_{j=1}^{K-1}z_j \mathds 1_{\{ \mathcal V^{-j}_{t-} \le \xi \}} N^{-M}\left(\d t, \d \xi, \prod_{k=1}^{K-1}\d z_k\right)\\
    &+ \int_{z \in \mathbb R_+} \left(S^1_{t-} - z - S^{-K}_{t-} \right) \mathds 1_{\left\{ S^1_{t-} - z \in \left[ S^{-K}_{t-},  S^{-K+1}_{t-}\right[ \right\}} N^{-L}(\d t, \d z)\\
    &+  \int_{z \in \mathbb R_+} \left( S^{-K+1}_{t-} - S^{-K}_{t-}\right) \mathds 1_{\left\{ S^1_{t-} - z > S^{-K+1}_{t-} \right\}} N^{-L}(\d t, \d z) - \sum_{l=1}^{K}  \int_{z \in \mathbb R_+} z N^{-l,C}(\d t, \d z),
\end{align*}
and similarly
\begin{align*}
    \d S^{K}_t =&\int_{\xi \in \mathbb R_+} \int_{(z_1, \ldots, z_{K-1}) \in \mathbb R^{K-1}_+ } \sum_{j=1}^{K-1}z_j \mathds 1_{\{ \mathcal V^{j}_{t-} \le \xi \}} N^{M}\left(\d t, \d \xi, \prod_{k=1}^{K-1}\d z_k\right)\\
    &+ \int_{z \in \mathbb R_+} \left(S^{-1}_{t-} + z - S^K_t \right) \mathds 1_{\left\{ S^{-1}_{t-} + z \in \left] S^{K-1}_{t-},  S^{K}_{t-}\right]\right\}} N^{L}(\d t, \d z)\\
    &+\int_{z \in \mathbb R_+} \left( S^{K-1}_{t-} - S^{K}_{t-}\right) \mathds 1_{\left\{ S^{-1}_{t-} + z < S^{K-1}_{t-} \right\}} N^{L}(\d t, \d z) + \sum_{l=1}^K \int_{z \in \mathbb R_+} z N^{l,C}(\d t, \d z).
\end{align*}
In both these equations, the first term corresponds to the arrival of a market order. In this case, as soon as the size of the market order exceeds the side of the first limit, the $K$-th limit has to be replaced by a new limit in the form of a random variable.\\

The second term corresponds to the arrival of a limit order between the $(K-1)$-th and the $K$-th limit: in that case, the new $K$-th limit is this new limit order. The third term deals with limit orders arriving before the $(K-1)$-th limit, in which case the $K$-th limit is simply shifted: the new $K$-th limit is the previous $(K-1)$-th limit.\\

The last term corresponds to the cancellation of orders: in that case, the new $K$-th has to be introduced in the form of a random variable. 

\subsubsection{Volumes dynamics}

We can now write the dynamics of $(\bar V_t)_t$. For $k=1$, we have
\begin{align*}
    \d V^{-1}_t =&\int_{\xi \in \mathbb R_+} \left(\sum_{i=1}^{K-1} \left( \mathcal V^{-i-1}_{t-} -  \mathcal V^{-i}_{t-} \right) \mathds{1}_{\{ \mathcal V^{-i}_{t-} \le \xi \}} - \xi \right) N^{-M}(\d t, \d \xi) + \left(V^{-2}_{t-} - V^{-1}_{t-} \right) \d N^{-1,C}_t\\
    & + \int_{z \in \mathbb R_+} \int_{\zeta \in \mathbb R_+} \left(\zeta - V^{-1}_{t-} \right) \mathds 1_{\left\{ S^1_{t-} - z \in \left[ S^{-1}_{t-},  S^{1}_{t-}\right[ \right\}} N^{-L}(\d t, \d z, \d \zeta),
\end{align*}
and similarly
\begin{align*}
    \d V^{1}_t =&\int_{\xi \in \mathbb R_+} \left( \sum_{i=1}^{K-1 }\left( \mathcal V^{i+1}_{t-} - \mathcal V^{i}_{t-} \right)\mathds{1}_{\{ \mathcal V^{i}_{t-} \le \xi \}} -\xi \right)N^{M}(\d t, \d \xi) + \left(V^{2}_{t-} - V^{1}_{t-} \right) \d N^{1,C}_t\\
    &+  \int_{z \in \mathbb R_+} \int_{\zeta \in \mathbb R_+} \left(\zeta - V^1_t \right) \mathds 1_{\left\{ S^{-1}_{t-} + z \in \left] S^{-1}_{t-},  S^{1}_{t-}\right]\right\}} N^{L}(\d t, \d z, \d \zeta).
\end{align*}
In both these equations, the first term corresponds to the arrival of a market order. If the size of the market order exceeds the cumulative volume of the first $i$ limits but not that of the first $i+1$ limits, then the new first limit is the previous $i+1$-th limit, and the corresponding volume is the previous the cumulative volume of the first $i+1$ limits minus the volume of the market order. The second term corresponds to the cancellation of the order at the first limit: in that case, the new first limit is the previous second limit, and the volumes are shifted accordingly. Cancellation of orders at other limits do not impact the first limit. Finally, the last term corresponds to the arrival of a limit order within the current bid-ask spread: in that case, the new first limit is this new limit order, and the corresponding volume is introduced in the form of a random variable.\\

For $1<k<K$, we have
\begin{align*}
    \d V^{-k}_t =&\int_{\xi \in \mathbb R_+} \int_{(z_1, \ldots, z_{K-1}) \in \mathbb R^{K-1}_+ } \int_{(\zeta_1, \ldots, \zeta_{K-1}) \in \mathbb R^{K-1}_+ } \Bigg( \mathds{1}_{\{  \xi \ge \mathcal V^{-K+k-1}_{t-} \}} \bigg(\zeta_{K-k+1} -V^{-k}_{t-}\bigg)\\
    &\qquad+ \mathds{1}_{\{ \xi < \mathcal V^{-K+k-1}_{t-} \}} \Big(\sum_{i=1}^{K-k} \left(\mathcal V^{-k-i}_{t-} - \mathcal V^{-k-i+1}_{t-} \right) \mathds{1}_{\{ \mathcal V^{-i}_{t-} \le \xi  \}} - \xi \Big) \Bigg) N^{-M}\left(\d t, \d \xi, \prod_{k=1}^{K-1}\d z_k, \prod_{k=1}^{K-1}\d \zeta_k\right)\\
    &+ \int_{z \in \mathbb R_+} \int_{\zeta \in \mathbb R_+} \left(\zeta - V^{-k}_{t-} \right) \mathds 1_{\left\{ S^1_{t-} - z \in \left[ S^{-k}_{t-},  S^{-k+1}_{t-}\right[ \right\}} N^{-L}(\d t, \d z, \d \zeta)\\
    &+  \int_{z \in \mathbb R_+} \left( V^{-k+1}_{t-} - V^{-k}_{t-}\right) \mathds 1_{\left\{ S^1_{t-} - z > S^{-k+1}_{t-} \right\}} N^{-L}(\d t, \d z) + \sum_{l=1}^k \left(V^{-k-1}_{t-} - V^{-k}_{t-} \right) \d N^{-l,C}_t,
\end{align*}
and similarly
\begin{align*}
    \d V^{k}_t =&\int_{\xi \in \mathbb R_+} \int_{(z_1, \ldots, z_{K-1}) \in \mathbb R^{K-1}_+ } \int_{(\zeta_1, \ldots, \zeta_{K-1}) \in \mathbb R^{K-1}_+ } \Bigg( \mathds{1}_{\{  \xi \ge \mathcal V^{K-k+1}_{t-} \}} \bigg( \zeta_{K-k+1} -V^{k}_{t-}\bigg)\\
    &\qquad+ \mathds{1}_{\{ \xi < \mathcal V^{K-k+1}_{t-} \}}\Big(\sum_{i=1}^{K-k} \left(\mathcal V^{k+i}_{t-} - \mathcal V^{k+i-1}_{t-} \right) \mathds{1}_{\{ \mathcal V^{i}_{t-} \le \xi  \}} - \xi \Big) \Bigg) N^{M}\left(\d t, \d \xi, \prod_{k=1}^{K-1}\d z_k, \prod_{k=1}^{K-1}\d \zeta_k\right)\\
    & + \int_{z \in \mathbb R_+} \int_{\zeta \in \mathbb R_+}  \left(\zeta  - V^k_t \right) \mathds 1_{\left\{ S^{-1}_{t-} + z \in \left] S^{k-1}_{t-},  S^{k}_{t-}\right]\right\}} N^{L}(\d t, \d z, \d \zeta)\\
    &+\int_{z \in \mathbb R_+} \left( V^{k-1}_{t-} - V^{k}_{t-}\right) \mathds 1_{\left\{ S^{-1}_{t-} + z < S^{k-1}_t \right\}} N^{L}(\d t, \d z) + \sum_{l=1}^k \left(V^{k+1}_{t-} - V^{k}_{t-} \right) \d N^{l,C}_t.
\end{align*}
In both these equations, the first term corresponds to the arrival of a market order. It can be decomposed into two parts: the first part deals with the case where the market order consumes $K-k+1$ limits (or more), in which case new limits have to be introduce after the (previous) last limit, and the corresponding volumes are introduced in the form of random variables; the second part deals with the case where the market order only consumes a small number of limits, so that the $k$-th limit can be replaced by an already existing limit and does not have to be drawn as a random variable, and the volumes are then shifted accordingly.\\

The second term corresponds to the arrival of a limit order between the $(k-1)$-th and the $k$-th limit: in that case, the new $k$-th limit is this new limit order, and the corresponding volume is introduced in the form of a random variable. The third term deals with limit orders arriving before the $(k-1)$-th limit, in which case the $k$-th limit is simply shifted.\\

The last term corresponds to the cancellation of orders before the $k$-th limit: in that case, the new $k$-th limit is the previous $(k+1)$-th limit, and volumes are just shifted accordingly. Cancellation of orders at limits above $k$ do not impact the $k$-th limit. \\

Finally, for $k=K$ we have
\begin{align*}
    \d V^{-K}_t =&\int_{\xi \in \mathbb R_+} \int_{(z_1, \ldots, z_{K-1}) \in \mathbb R^{K-1}_+ } \int_{(\zeta_1, \ldots, \zeta_{K-1}) \in \mathbb R^{K-1}_+ } \mathds 1_{\{ \mathcal V^{-1}_{t-} \le \xi \}}\left(\zeta_1 - V^{-K}_{t-} \right)  N^{-M}\left(\d t, \d \xi, \prod_{k=1}^{K-1}\d z_k, \prod_{k=1}^{K-1}\d \zeta_k\right)\\
    &+ \int_{z \in \mathbb R_+} \int_{\zeta \in \mathbb R_+} \left(\zeta - V^{-K}_{t-} \right) \mathds 1_{\left\{ S^1_{t-} - z \in \left[ S^{-K}_{t-},  S^{-K+1}_{t-}\right[ \right\}} N^{-L}(\d t, \d z, \d \zeta)\\
    &+  \int_{z \in \mathbb R_+} \left( V^{-K+1}_{t-} - V^{-K}_{t-}\right) \mathds 1_{\left\{ S^1_{t-} - z > S^{-K+1}_{t-} \right\}} N^{-L}(\d t, \d z) + \sum_{l=1}^{K}  \int_{z\in \mathbb R_+} \int_{\zeta \in \mathbb R_+} \left(\zeta - V^{-K}_{t-} \right) N^{-l,C}(\d t, \d z),
\end{align*}
and similarly
\begin{align*}
    \d V^{K}_t =&\int_{\xi \in \mathbb R_+} \int_{(z_1, \ldots, z_{K-1}) \in \mathbb R^{K-1}_+ } \int_{(\zeta_1, \ldots, \zeta_{K-1}) \in \mathbb R^{K-1}_+ } \mathds 1_{\{ \mathcal V^{1}_{t-} \le \xi \}} \left(\zeta_1 - V^{K}_{t-} \right)  N^{M}\left(\d t, \d \xi, \prod_{k=1}^{K-1}\d z_k, \prod_{k=1}^{K-1}\d \zeta_k\right)\\
    &+ \int_{z \in \mathbb R_+} \int_{\zeta \in \mathbb R_+} \left(\zeta - V^{K}_{t-} \right) \mathds 1_{\left\{ S^{-1}_{t-} + z \in \left[ S^{K-1}_{t-},  S^{K}_{t-}\right[ \right\}} N^{L}(\d t, \d z, \d \zeta)\\
    &+\int_{z \in \mathbb R_+} \left( V^{K-1}_{t-} - V^{K}_{t-}\right) \mathds 1_{\left\{ S^{-1}_{t-} + z < S^{K-1}_t{t-}\right\}} N^{L}(\d t, \d z) + \sum_{l=1}^{K}  \int_{z\in \mathbb R_+} \int_{\zeta \in \mathbb R_+} \left(\zeta - V^{K}_{t-} \right) N^{l,C}(\d t, \d z).
\end{align*}
In both these equations, the first term corresponds to the arrival of a market order. In this case, as soon as the size of the market order exceeds the side of the first limit, the $K$-th limit has to be replaced by a new limit in the form of a random variable, and the corresponding volume is drawn as well.\\

The second term corresponds to the arrival of a limit order between the $(K-1)$-th and the $K$-th limit: in that case, the new $K$-th limit is this new limit order, and the volume is introduced in the ofrm of a random variable. The third term deals with limit orders arriving before the $(K-1)$-th limit, in which case the $K$-th limit is simply shifted: the new $K$-th limit is the previous $(K-1)$-th limit, and the volumes are shifted accordingly.\\

The last term corresponds to the cancellation of orders: in that case, the new $K$-th has to be introduced in the form of a random variable, and the corresponding volume as well.

\section{Numerical analysis}\label{sec:num}

\subsection{Parameters}

\begin{figure}
    \centering
    \begin{subfigure}{.45\linewidth}
        \includegraphics[width=\linewidth]{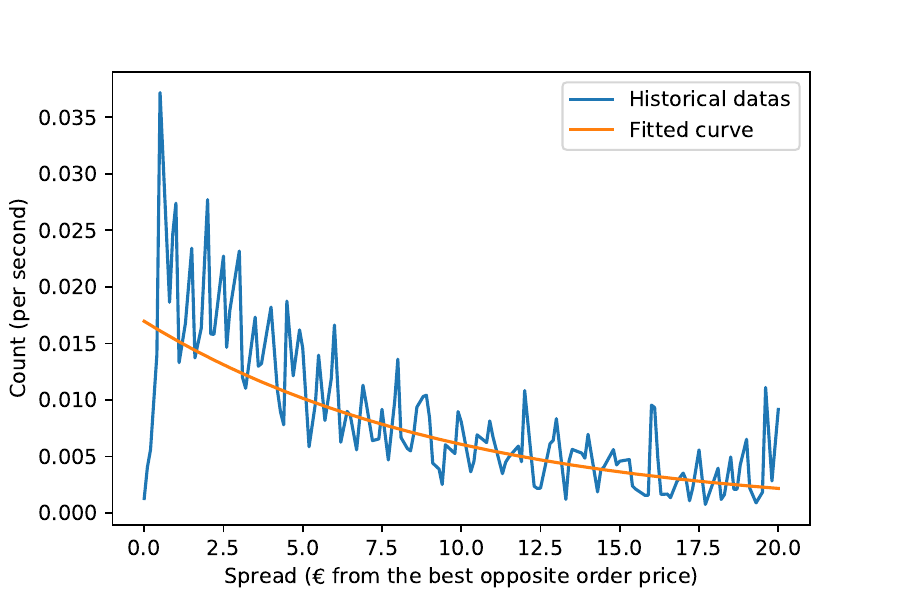}
        \caption{\label{fig:spread_MO} Spread dependence of market orders}
    \end{subfigure}
    \begin{subfigure}{.45\linewidth}
        \includegraphics[width=\linewidth]{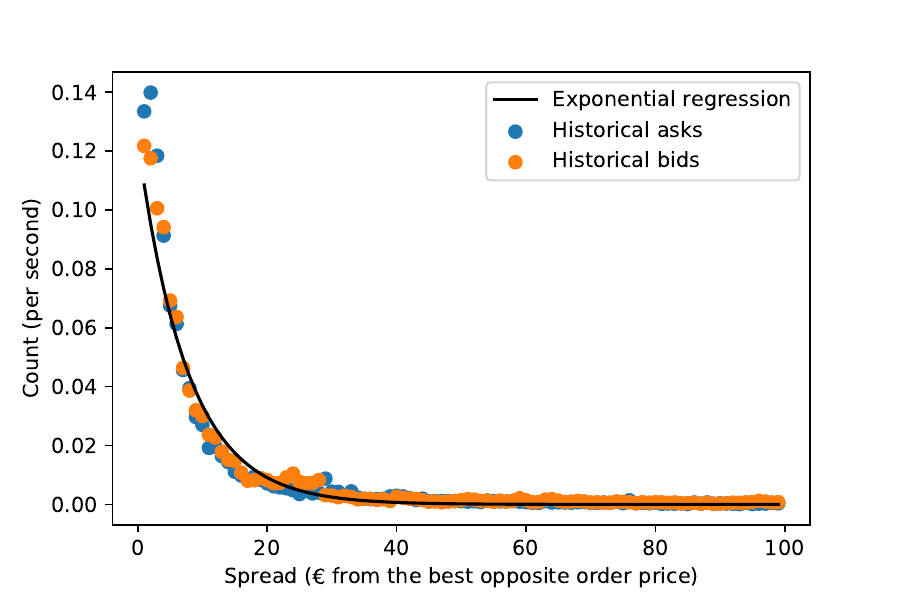}
        \caption{\label{fig:distrib_LO} New limit order distribution}
    \end{subfigure}
    \caption{Historical data approximation}
    \label{fig:hist_data}
\end{figure}

For our numerical experiments, we consider the last 4 hours (i.e. $T=4\;h$) of trading for the 6 p.m. maturity. \footnote{We focus exclusively on the XBID market, which closes one hour before the delivery period. Therefore, we analyze data from 1 p.m. to 5 p.m. for a delivery maturity at 6 p.m.} We chose to only represent the 5 first limits on each side of the order book, i.e. $K=5$. In this section, we describe the functional forms and parameters used for the simulations. The data suggest some spread dependencies and symmetric parameters at bid and ask sides. These can be seen in Figure \ref{fig:hist_data}. \\

For the market orders, the data suggest a function $\lambda^M = \lambda^{-M}$ of the form:
$$\lambda^M(t, \mathbf S)  = \lambda^{-M}(t, \mathbf S) = \bar \lambda^M e^{-\kappa^M(T-t)} e^{-\beta^M \mathbf S}$$

for the intensity of arrival, and we approximate the distribution $\phi^M(\xi, \mathcal V)\d \xi = \phi^{-M}(\xi, \mathcal V) \d \xi$ by a discrete distribution of the form
$$ \sum_{i=1}^6 p_i^M \delta_{v^M_i \wedge \mathcal V}(\d \xi).$$
 This distribution is bounded by $\mathcal V$ in accordance with our assumption that a single market order cannot consume the $K$ first limits of the order book.\\
 
Using standard statistical estimation methods such as least squares regression (see \citeauthor{laruellehors} \cite{laruellehors} for more details), we obtain the following values:
\begin{itemize}
    \item $\bar \lambda^M = 45.72 \;h^{-1}$
    \item $\kappa^M = 0.51 \;h^{-1}$
    \item $\beta^M = 0.5 \;\text{\euro}^{-1}$
    \item $p^M = (0.480, 0.158, 0.314, 0.032, 0.012, 0.004)$
    \item $v^M = (1., 2., 5., 10., 15., 25.) \;MWh.$
\end{itemize}

Similarly for the limit orders, the data suggest a function $\lambda^L = \lambda^{-L}$ of the form
$$\lambda^L(t)  = \lambda^{-L}(t) = \bar \lambda^L e^{-\kappa^L(T-t)} $$
for the intensity of arrival,  a probability density function
$$f^L(t, z) = f^{-L}(t, z) = \mathds 1_{z>0} \beta^L_t e^{-\beta^L_t z} $$
for the distance to the best price on the opposite side, corresponding to an exponential distribution with a time varying parameter given by $\beta^L_t = A^L e^{-b^L(T-t)}$. 
As before, we approximate the distribution $\phi^L(\zeta)\d \zeta = \phi^{-L}(\zeta) \d \zeta$ by a discrete distribution of the form
$$ \sum_{i=1}^6 p_i^L \delta_{v^L_i }(\d \zeta).$$
 
Again, a standard statistical estimation method gives the following values:
\begin{itemize}
    \item $\bar \lambda^L = 450 \;h^{-1}$
    \item $\kappa^L = 5.22 \cdot 10^{-4} \;h^{-1}$
    \item $A^L = 0.145\;\text{\euro}^{-1}$
    \item $b^L = 0.02\; h^{-1}$
    \item $p^L = (0.322, 0.152, 0.464, 0.022, 0.011, 0.029)$
    \item $v^L = (1., 2., 5., 10., 15., 25.) \;MWh.$
\end{itemize}

For the cancel orders, we use again a function $\lambda^C = \lambda^{-C}$ of the form
$$\lambda^C(t) = \lambda^{-C}(t) = \bar \lambda^C e^{-\kappa^C(T-t)},$$
with the following values
\begin{itemize}
    \item $\bar \lambda^C = 72\; h^{-1}$
    \item $\kappa^C = 0.6\; h^{-1}.$
\end{itemize}

Finally, for a newly inserted order after the $K-$th limit, it is natural to consider
$$f^{K+1}(t,z) = f^{-K-1}(t,z) = f^L(t,z) = \mathds 1_{z>0} \beta^L_t e^{-\beta^L_t z},$$
and similarly we approximate the distribution $\phi^{K+1}(\zeta)  \d \zeta = \phi^{-K-1}(\zeta)  \d \zeta$ by the discrete distribution 
$$ \sum_{i=1}^6 p_i^L \delta_{v^L_i }(\d \zeta).$$

\subsection{Results}\label{sec:Results}

In this section, we run different simulations of the order book using the parameters introduced above. and we analyze the results. We stop the simulations one hour before maturity, as it corresponds to the decoupling of the electricity markets. We start with an initial mid price of $50\;\text{\euro}$, and consider that the order book at the initial time is symmetric, with limits at the bid placed at the prices $45\;\text{\euro}, 44\;\text{\euro}, 42\;\text{\euro}, 39\;\text{\euro}, 35\;\text{\euro}$, and limits at the ask placed at the prices $55\;\text{\euro}, 56\;\text{\euro}, 58\;\text{\euro}, 61\;\text{\euro}, 65\;\text{\euro}$.\\

\begin{figure}[h!]
    \centering
    
    \begin{subfigure}{.45\linewidth}
      \includegraphics[width=\linewidth]{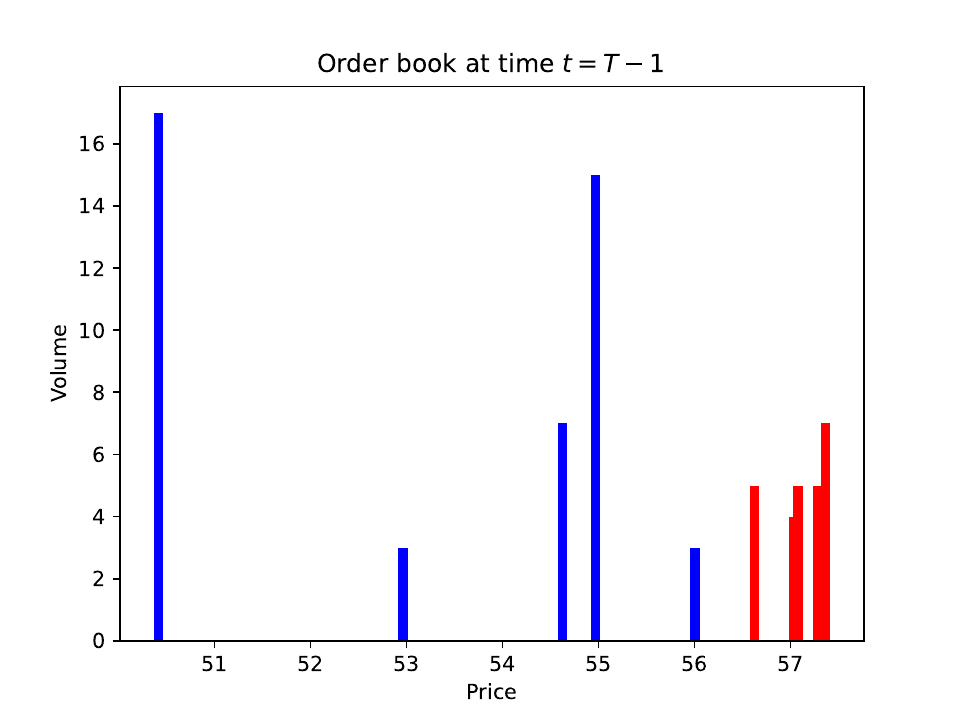}
      \caption{}
      \label{fob0}
    \end{subfigure}
    \begin{subfigure}{.45\linewidth}
      \includegraphics[width=\linewidth]{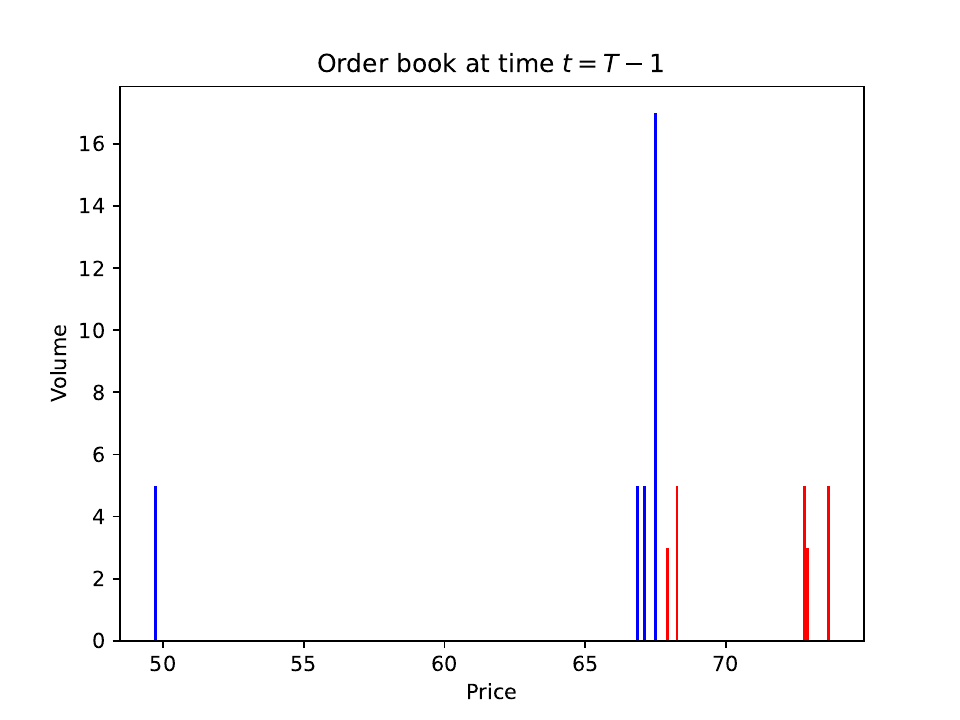}
      \caption{}
      \label{fob1}
    \end{subfigure}
    
    \medskip % create some *vertical* separation between the graphs
    \begin{subfigure}{.45\linewidth}
      \includegraphics[width=\linewidth]{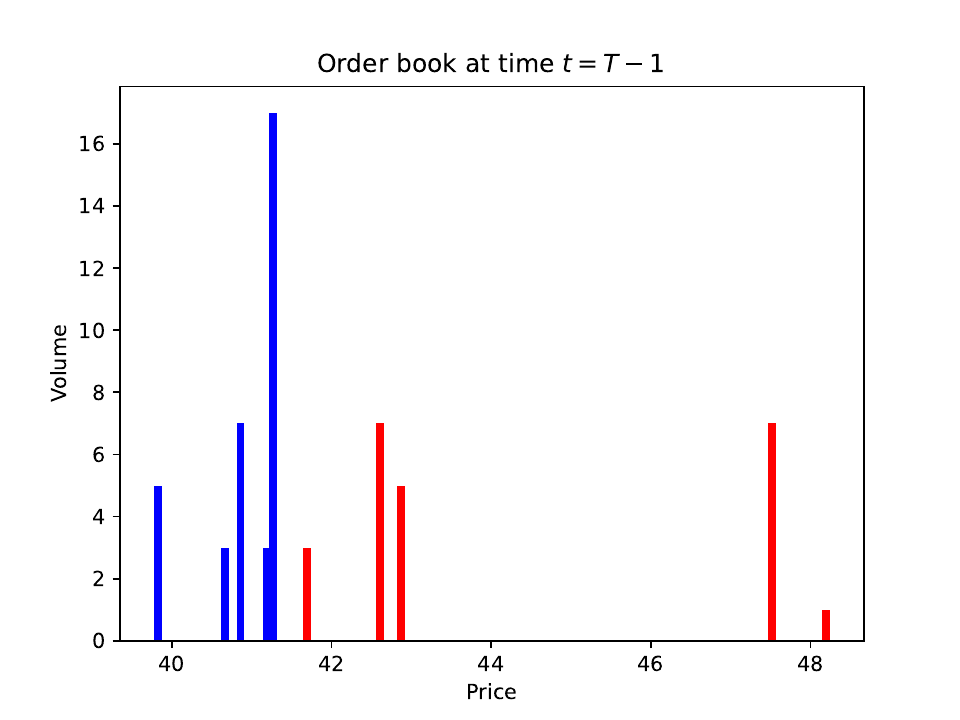}
      \caption{}
      \label{fob2}
    \end{subfigure}
    \begin{subfigure}{.45\linewidth}
      \includegraphics[width=\linewidth]{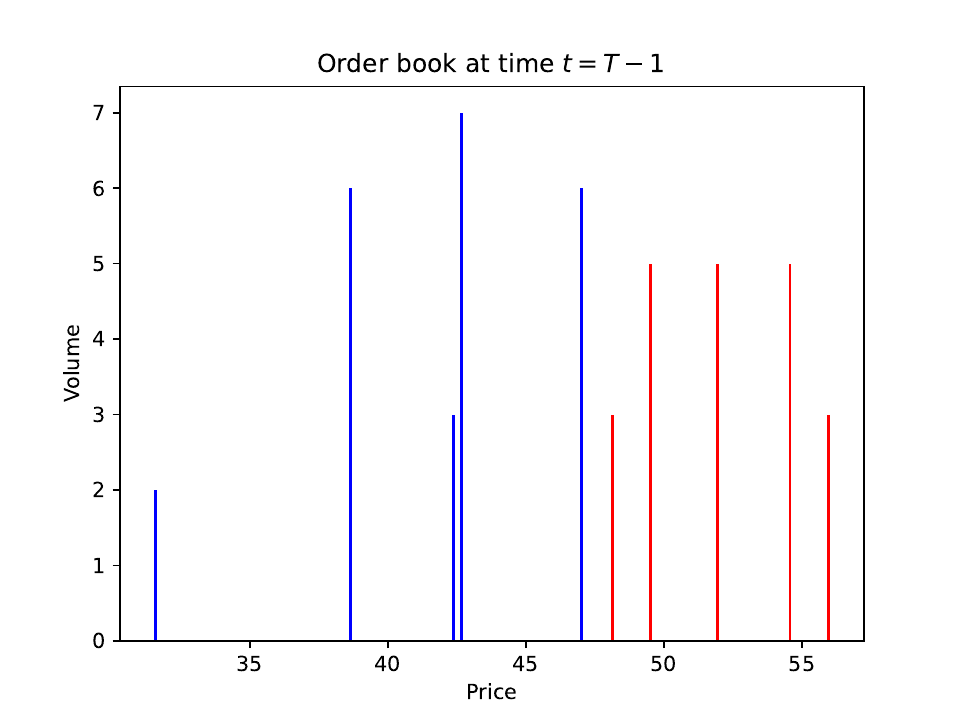}
      \caption{}
      \label{fob3}
    \end{subfigure}
    
    \caption{Simulations of the order book at $T-1$ (bid in blue, ask in red).}
    \label{fig:obsim}
\end{figure}

\begin{figure}[h!]
    \centering
    
    \begin{subfigure}{.45\linewidth}
      \includegraphics[width=\linewidth]{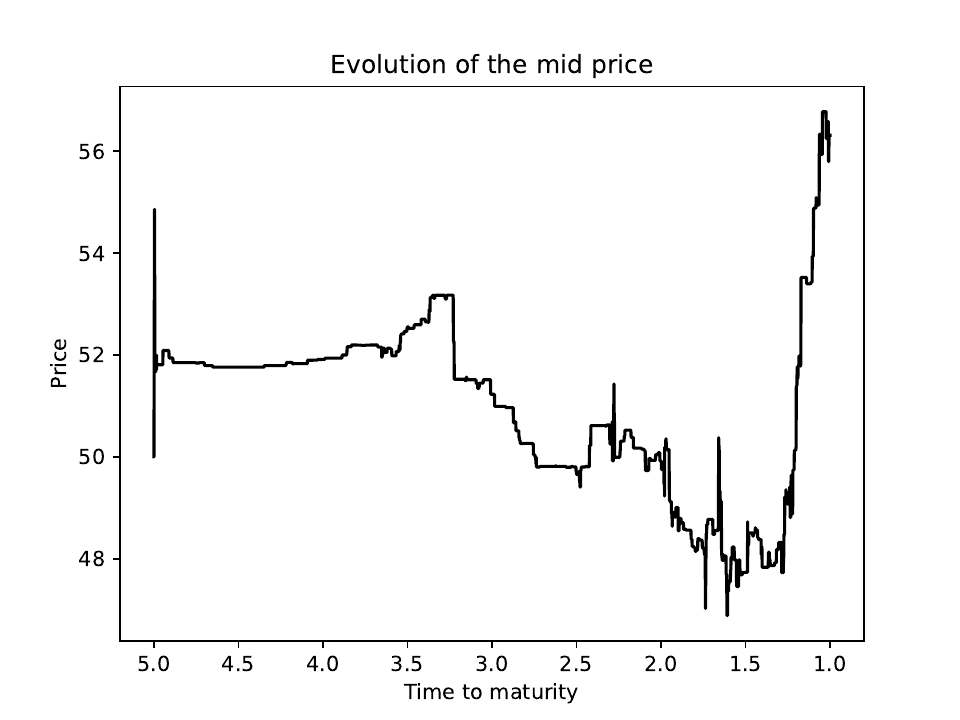}
      \caption{}
      \label{fob0}
    \end{subfigure}
    \begin{subfigure}{.45\linewidth}
      \includegraphics[width=\linewidth]{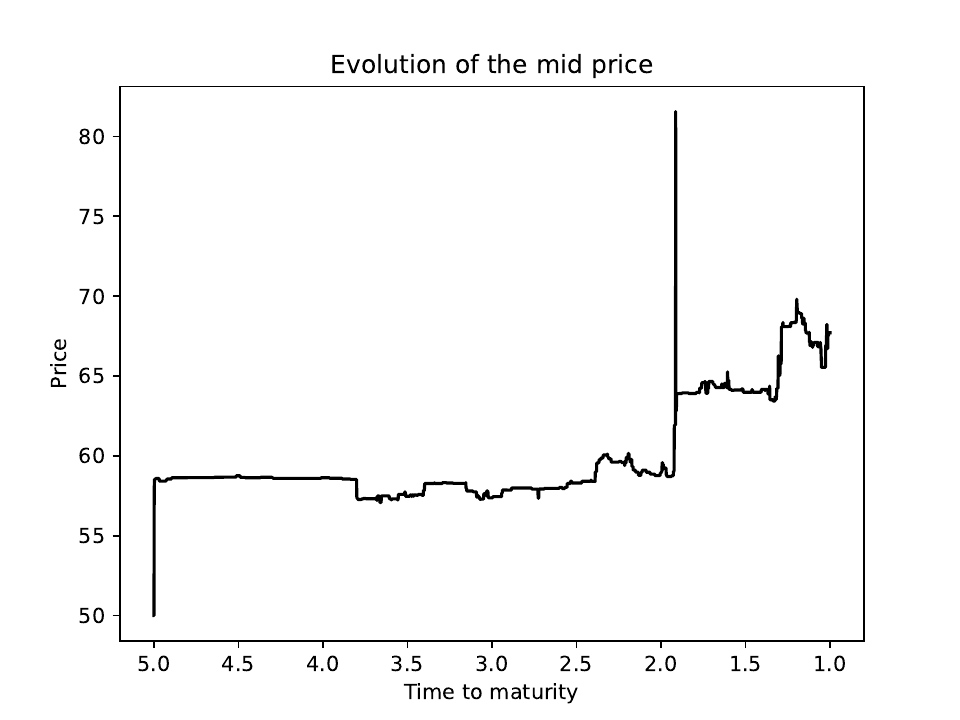}
      \caption{}
      \label{fob1}
    \end{subfigure}
    
    \medskip % create some *vertical* separation between the graphs
    \begin{subfigure}{.45\linewidth}
      \includegraphics[width=\linewidth]{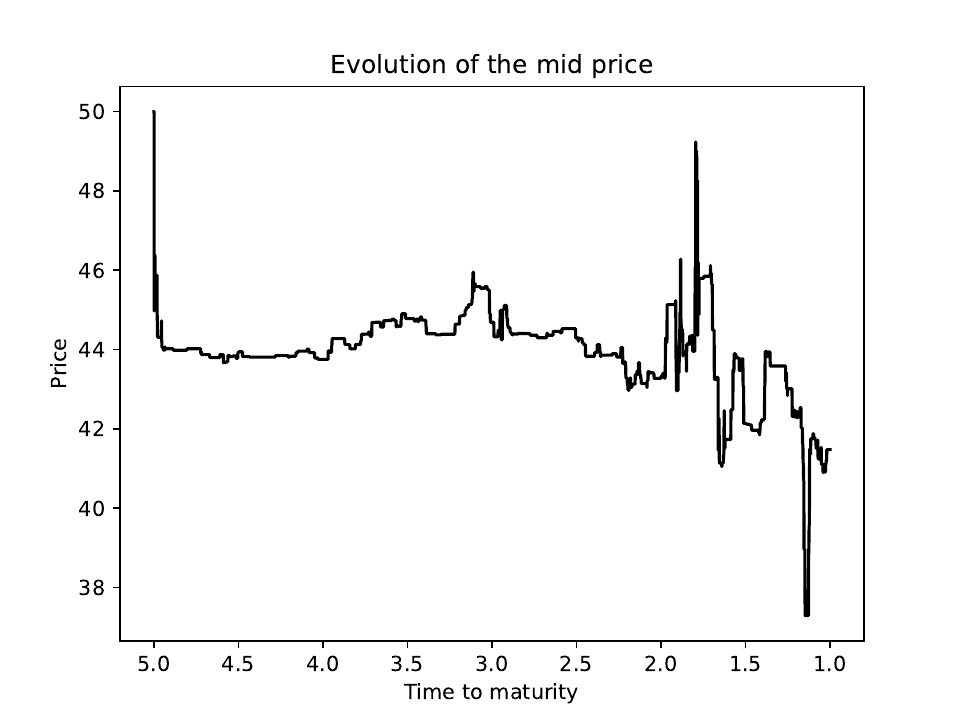}
      \caption{}
      \label{fob2}
    \end{subfigure}
    \begin{subfigure}{.45\linewidth}
      \includegraphics[width=\linewidth]{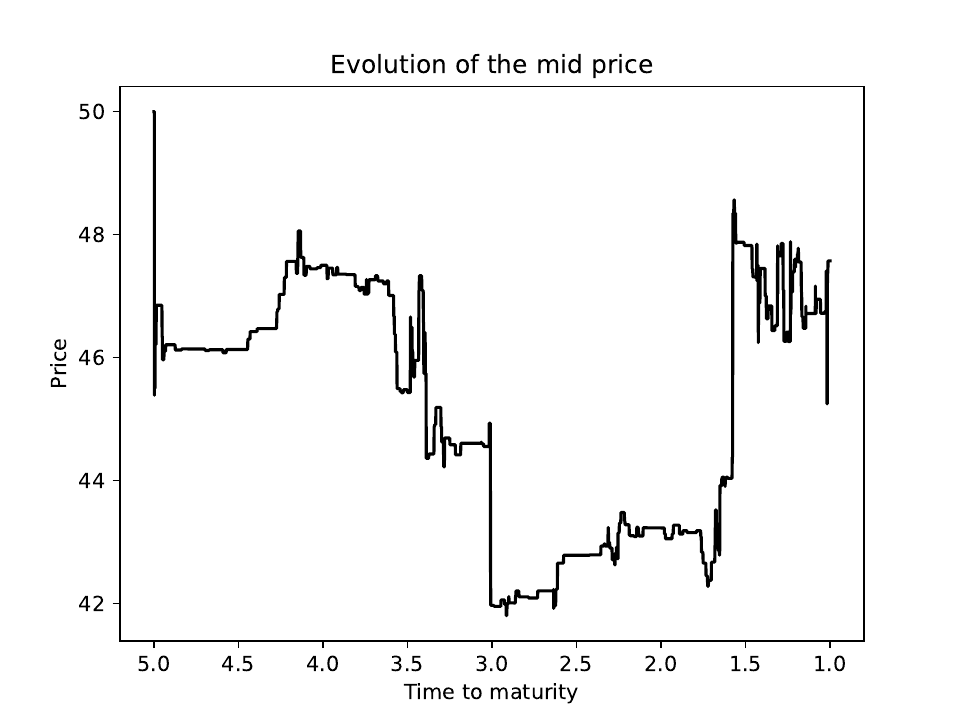}
      \caption{}
      \label{fob3}
    \end{subfigure}
    
    \caption{Simulations of the mid price}
    \label{fig:midpricesim}
\end{figure}

Figure \ref{fig:obsim} shows four simulations of the state of the order book one hour before maturity. We observe in particular that, as expected, our model allows for very large spreads between the different limits of the order book. Figure \ref{fig:midpricesim} shows the corresponding prices. We see that price movements increase when time gets close to maturity, which corresponds to what we observe in practice -- see Figure \ref{fig:midprice_evol}.

\begin{figure}

    \begin{subfigure}{.34\linewidth}
        \includegraphics[width=\linewidth]{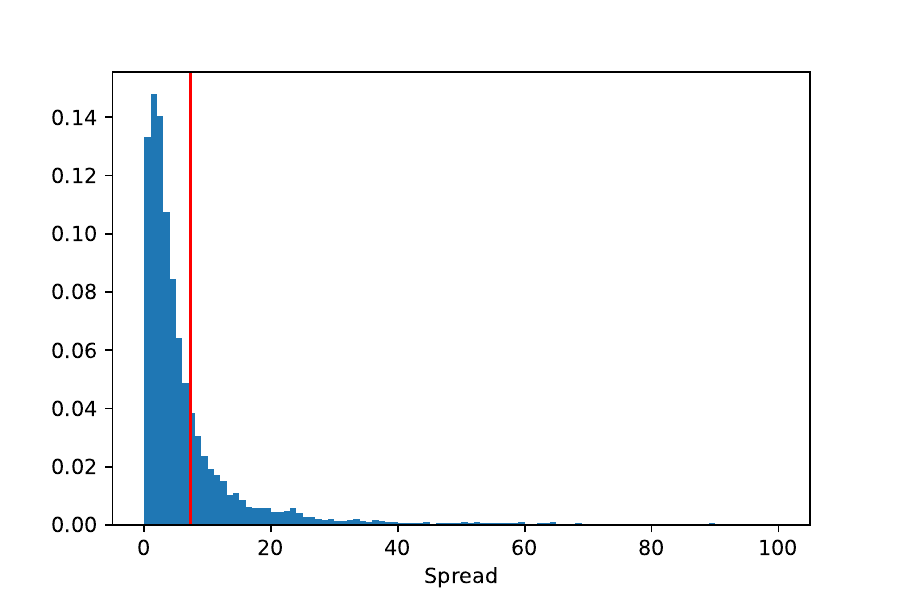}
        \caption{\label{fig:ref_spread_dist} First limit}
    \end{subfigure}
    \begin{subfigure}{.34\linewidth}
        \includegraphics[width=\linewidth]{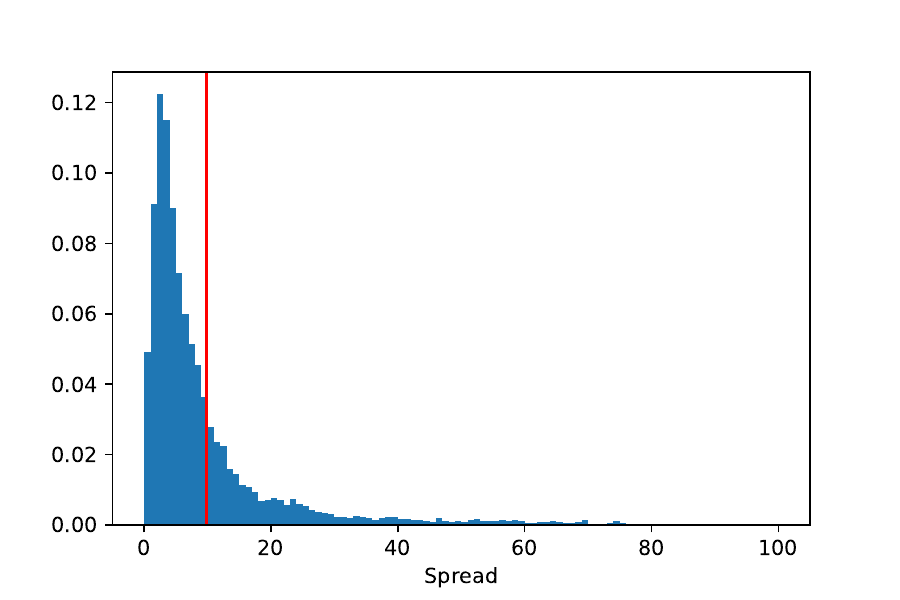}
        \caption{\label{fig:ref_2spread_dist} Second limit}
    \end{subfigure}
    \begin{subfigure}{.34\linewidth}
        \includegraphics[width=\linewidth]{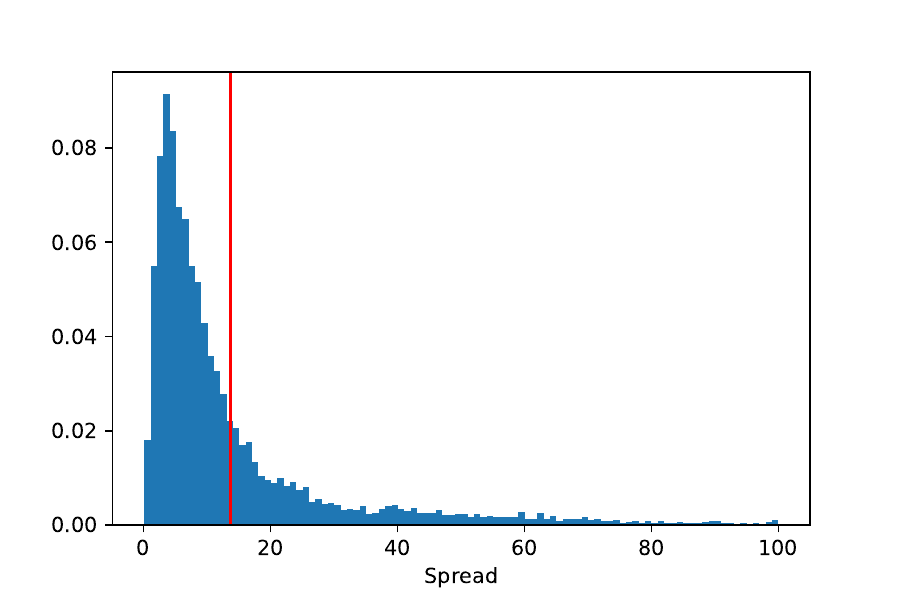}
        \caption{\label{fig:ref_3spread_dist} Third limit}
    \end{subfigure}

    \caption{\label{fig:ref_spread_distribution} Historical spread distribution of the first three limits on $t \in [T-5; T-1]$ on September 01, 2021 to November 30, 2021 (the red line represents the mean).}
\end{figure}

\begin{figure}

    \begin{subfigure}{.34\linewidth}
        \includegraphics[width=\linewidth]{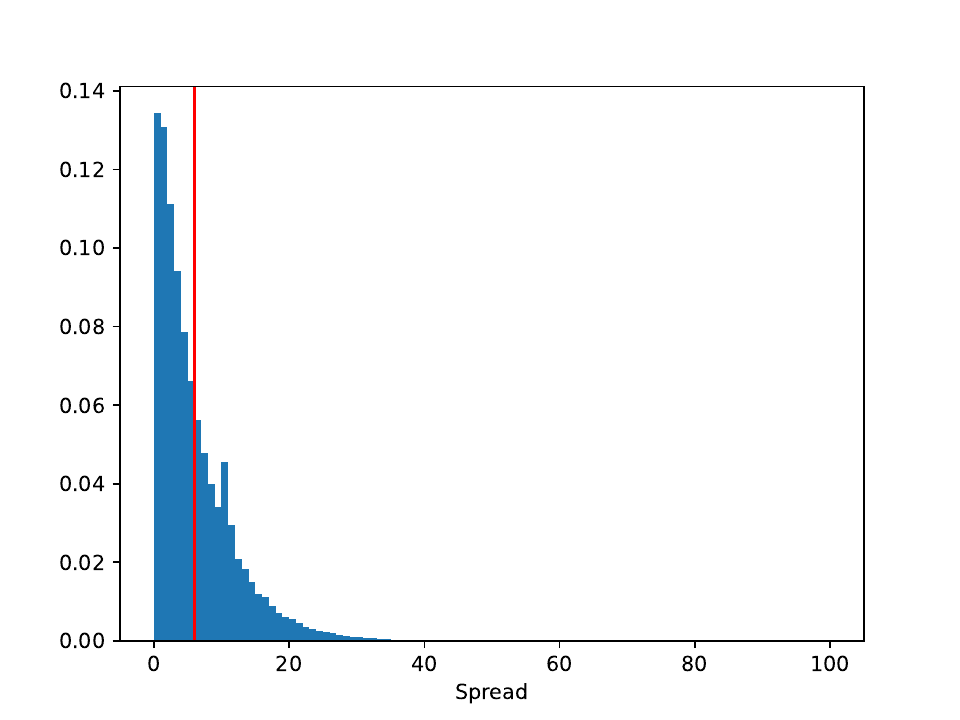}
        \caption{\label{fig:spread_dist} First limit}
    \end{subfigure}
    \begin{subfigure}{.34\linewidth}
        \includegraphics[width=\linewidth]{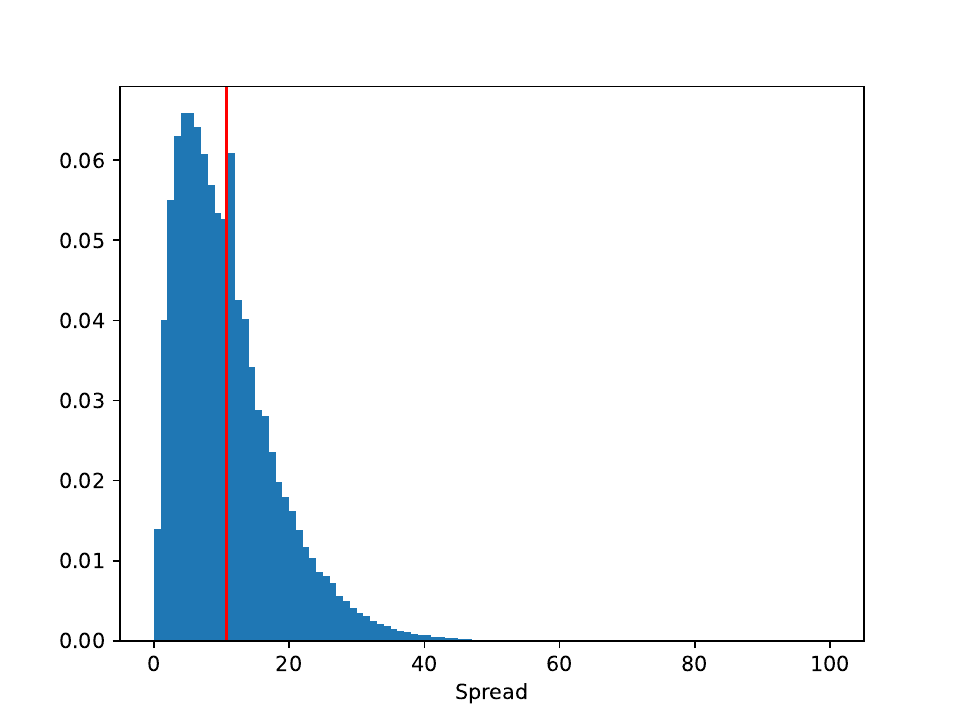}
        \caption{\label{fig:2spread_dist} Second limit}
    \end{subfigure}
    \begin{subfigure}{.34\linewidth}
        \includegraphics[width=\linewidth]{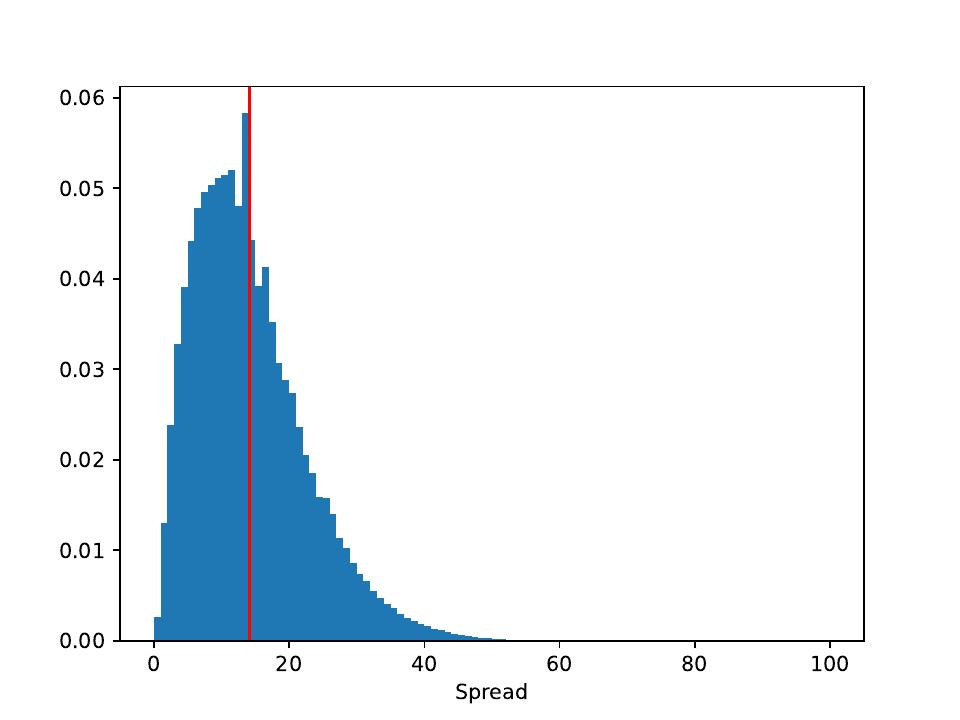}
        \caption{\label{fig:3spread_dist} Third limit}
    \end{subfigure}

    \caption{\label{fig:spread_distribution} Simulated spread distributions for 10000 simulations (the red line represents the mean).}
\end{figure}

Figure \ref{fig:spread_dist} shows the distribution of spreads one hour before maturity obtained with 10000 Monte-Carlo simulations. The average spread is around $6.10\; \text{\euro}$, which is in line with what we observe in practice (see figure \ref{fig:ref_spread_dist}) with an average spread around $7.30\; \text{\euro}$.\\

Similarly, we show in Figures \ref{fig:2spread_dist} and \ref{fig:3spread_dist} the distribution of the distance between the second limits on each side and between the third limits on each side, respectively. For the former, we obtain an average spread of $10.70\;\text{\euro}$, while for the latter we obtain an average spread of $14.20\;\text{\euro}$, which is again close to what we observe in practice as we can see in figures \ref{fig:ref_2spread_dist} and \ref{fig:ref_3spread_dist} with for the second limit an average price around $9.90\;\text{\euro}$, and on the third limit we have an average spread of $13.74\;\text{\euro}$. \footnote{The spike observed in Figures \ref{fig:2spread_dist} and \ref{fig:3spread_dist} near the mean of the distribution is due to our choice for the initial state of the order book; it disappears if one randomizes this initial condition.}\\

As expected, the main drawback of our Poisson model is the lack of extreme events: the simulated distribution always seem more localized around the mean than that of the historical data, the latter consistently exhibiting a slightly higher standard deviation. This can be solved for instance by incorporating Hawkes processes in the model.\\

\subsection{Microstructure noise}
\begin{figure}
    \centering
    \begin{subfigure}{.45\linewidth}
        \includegraphics[width=\linewidth]{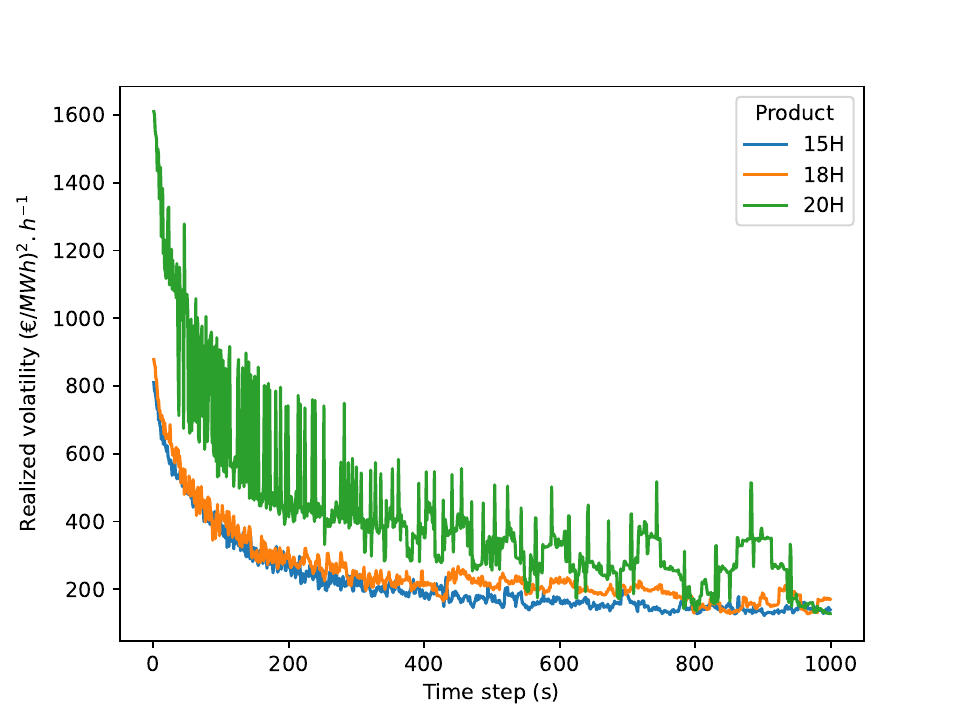}
        \caption{\label{fig:volatility_hist}Historical results on products 15H, 18H, 20H}
    \end{subfigure}
    \begin{subfigure}{.50\linewidth}
        \includegraphics[width=\linewidth]{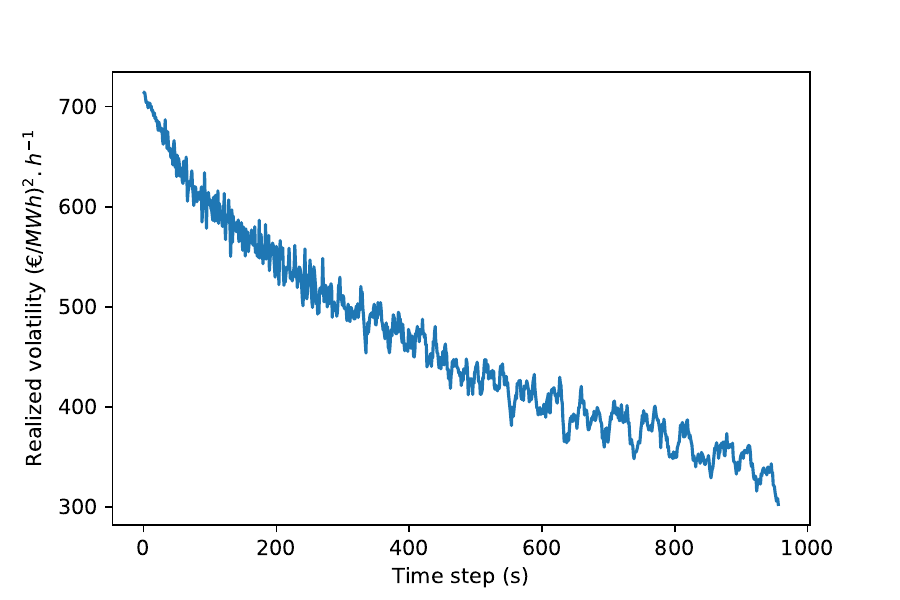}
        \caption{\label{fig:volatility_sim}Simulated results for a 18H calibration}
    \end{subfigure}
    \caption{Mean daily signature plots of the last 5 hours before the maturity computed and calibrated on all sessions between July and October, 2021}
    \label{fig:volatility}
\end{figure}

One of the main issue to reproduce LOBs is to take into account microstructure noise identified on electricity market by \citeauthor{deschatre2022electricity} \cite{deschatre2022electricity} and \citeauthor{deschatre2023common} \cite{deschatre2023common}. In this paper we use the empirical quadratic variation of the price process as a metric to assess the precision of our model. Figures \ref{fig:volatility} demonstrates that the signature plot of simulated LOB (using the average of 50 trajectories) is very similar to the empirical signature plots, computed from the quadratic variation of the mid-price $S(t)$ over a time period $[t_0,T]$ using a time step $\tau > 0$ where $t_0$ and $T$ respectively represent five hours and one hour before the maturity. The realized volatility function used here is: 

$$ \hat{C}(\tau) = \frac{1}{T}\sum_{n=0}^{T/\tau} |S((n+1)\tau) - S(n\tau)|^2 .$$

As anticipated, with an increase in the step time $\tau$, the realized volatility decreases until it stabilizes.\\

\clearpage
\bibliographystyle{plainnat}
\bibliography{biblio.bib}

\end{document}